\documentstyle[12pt]{article}

\newcommand{\ket}[1]{\left| {#1} \right>}
\newcommand{\bra}[1]{\left< {#1} \right|}

\newcommand{\beq}{\begin{eqnarray}}
\newcommand{\eeq}{\end{eqnarray}}

\begin{document}

\title{The Ion Trap Quantum Information Processor}
\author{Andrew Steane\\
Department of Physics, University of Oxford\\
Clarendon Laboratory, Parks Road, Oxford, OX1 3PU, England.}
\date{8 August 1996}
\maketitle
\begin{abstract}
An introductory review of the linear ion trap is given, with particular regard 
to its use for quantum information processing. The discussion aims to bring 
together ideas from information theory and experimental ion trapping, to 
provide a resource to workers unfamiliar with one or the other of these 
subjects. It is shown that information theory provides valuable concepts
for the experimental use of ion traps, especially error correction, and
conversely the ion trap provides a valuable link between information
theory and physics, with attendant physical insights. Example
parameters are given for the case of calcium ions. Passive
stabilisation will allow about 200 computing
operations on 10 ions; with error correction this can be greatly extended.
\end{abstract}

\section{Introduction}

This paper is a contribution to the rapidly developing field of quantum 
information theory and experiment. Quantum information is an interdisciplinary 
subject, in which computer scientists and other experts in the theory of 
classical information and computing are not necessarily familiar with quantum 
mechanics, while physicists and other experts in quantum theory are not 
necessarily familiar with information theory. Furthermore, whereas the field 
has enjoyed a rich theoretical treatment, there is a lack of an experimental 
basis to underpin the ideas. This is especially significant to the issue
of error correction, or more generally any stabilisation of a quantum
computer, which is among the most important unresolved issues in this field.
The aim of this paper is to offer an aid to people from different sides
of the subject to understand issues in the other. That is to say, the
ideas of quantum information and computing will be introduced to
experimental physicists, and a particular physical system which
might implement quantum computing will be described in detail for the
benefit of theoreticians. I hope to give sufficient information to
form more or less a `blueprint' for the type of quantum information
processor which is currently achievable in the lab, highlighting
the various experimental problems involved. The discussion is
like a review in that it brings together the work of other authors
rather than provides much original material. However, an exhaustive
review of the wide range of subjects involved is not intended, and as
a result it will not be possible to do justice to the efforts
of the many people who brought the experimental and theoretical
programmes to their present state of accomplishment.

The plan of the article is as follows. In section \ref{s:req}
the concepts of quantum information processing are introduced,
especially the `universal' set
of quantum logic gates. In section \ref{s:trap} the linear ion trap is
considered as a realisable system in which these ideas can be applied.
A physical process by which quantum logic gates may be applied in an
ion trap is described in detail. Limitations on the size of the
processor (number of quantum bits) and speed of operation
(`switching rate') are discussed. In sections \ref{s:cool} and \ref{s:rf}
the main experimental techniques required to realise the ion trap processor
in the lab are discussed; these are laser cooling of the ions, and low-noise
generation of the correct dc and radio frequency (rf) voltages for the trap 
electrodes, as well as a good choice of electrode design. In section 
\ref{s:ions} we begin to establish definite values for the experimental 
parameters, by considering specific candidate ions to which the methods can be 
applied. Example values are given for the singly-charged calcium ion. In 
section \ref{s:perf} experimental limitations such as unwanted 
heating of the ion motion are discussed.
This leads to an estimate for the maximum 
number of unitary operations (quantum gates) which could be carried out in 
the processor before the coherence of the system is destroyed. 
It is found that for an example case of around 10 ions, a few hundred
operations represents a severe experimental challenge. The
use of quantum error correction
to enhance the performance is then discussed. This
should allow great increases in the number of operations,
while preserving coherent evolution. The 
conclusion outlines the most important avenues for future investigation. 

\section{Requirements for Quantum Information Processing}
\label{s:req}

Quantum information theory is concerned with understanding the properties of 
quantum mechanics from an information theoretic point of view. This turns out 
to be a very fruitful approach, and leads naturally to the idea of information 
processing or computing, so that one poses the question ``what are the 
possibilities for, and the limitations of, information processing in a physical 
system governed by the laws of quantum mechanics?'' A great deal of theoretical 
insight into this question has been gained. For instance, it is possible to 
identify a small set of `building blocks' which if they could be realised and 
many of them combined, a `universal quantum computer' could be constructed. The 
computer is `universal' in the sense that it could simulate, by its 
computations, the action of any other computer, and so is more or less equal to 
or better than any other computer \cite{Turing}.
The phrase `more or less equal' has a 
technical definition which will be elaborated in
section \ref{s:effic}. A specific set of such 
building blocks is a set of two-state systems  (think of a line of spins), and 
a simple unitary interaction which can be applied at will to any chosen small 
set of these two-state systems \cite{Deutsch,qcomp}. In this context it is 
useful to describe the interaction in terms of its propagator $U = \exp(i H 
\Delta t / \hbar)$ rather than its Hamiltonian $H$. Here $\Delta t$ is some 
finite interval of time (one `clock period' in computing terminology) at the 
end of which the propagator has had just the effect desired on the computer. 
After this time the interaction $H$ falls to zero (is turned off). Such a 
propagator is referred to as a `gate', by analogy
with a logic gate in a classical computer. 

To do quantum information processing, these requirements may be summed up as 
that you need a system (`quantum computer' or QC) with a Hilbert space of 
sufficient number $D$ of dimensions, over which you have complete experimental 
control. That is, you can tell your system to go from any of its states to any 
other, without uncontrollable error processes such as relaxation and 
decoherence.

It is usual to consider a Hilbert space whose number of dimensions is a power 
of 2, ie $D=2^K$, in which case we say we have a system of $K$ quantum bits or 
`qubits'. Examples of qubits are the spin state of an
electron (2 orthogonal states and so a single qubit)
the polarization state of a photon (a single qubit), the internal state
of an atom having two energy levels of total spin $1$ and $2$ (8 states
and so 3 qubits). Whereas these are all equivalent from the 
point of view of the properties of Hilbert space, they are very different from 
the point of view of experimental implementation. The use of the word `qubit' 
rather than `two-state system' emphasizes this equivalence between otherwise 
very difference quantum systems. In fact, the idea of a qubit has further 
significance, since it can be shown \cite{Joz94,Schu95} that the essential 
properties of any quantum state of any system can transposed (by interactions 
allowed by the laws of physics) into the properties of a finite set of qubits 
and back again \cite{CD96}. The important point is that the average number of 
qubits required to do this is equal to the von Neumann entropy of the initial 
state (``quantum noiseless coding theorem'', also referred to as ``quantum data 
compression'' \cite{BenPT95}). Therefore the qubit gives a measure of {\em 
information} content in quantum systems, and is thus the correct quantum 
equivalent of the classical bit. 

Having accepted the invention of a new word for the quantum two-state system, 
there is justified resistance to the adoption of the terms `computer' and 
`computing' to describe the larger quantum systems with which we are concerned. 
This is because it is an open question whether a true quantum `computer' could 
ever function, since once the physical system has sufficient degrees of freedom 
to be meaningfully called a `computer', the large-scale interference necessary 
for parallel quantum computing may always be destroyed in practice, owing to 
the sensitivity of such interference to decoherence. For this 
reason, the more modest term `information processor' is used here as much as 
possible. The `processing' might consist of quite simple manipulations, such as 
allowing one qubit to interact with another, followed by a measurement of the 
state of the second qubit. Even such a simple operation has a practical use, 
since it can be used for error detection at the receiving end of a quantum 
communication channel, leading to the possibility of secure quantum key 
distribution for cryptography \cite{QPA,Benn95,Benn96}. 

Decoherence and dissipation in quantum mechanics is a subject in its own 
right, and has been discussed since the birth of quantum theory. Recent 
reviews and references may be found in \cite{Car93,Cal85,Zur}. Its impact on 
quantum computers in particular has been considered 
\cite{Land,Unruh95,Palma}, and will be taken into account in section 
\ref{s:perf}. 

It can be shown \cite{ugate} that to produce arbitrary unitary transformations 
of the state of a set of qubits, which is what one wants for information 
processing, it is sufficient to be able to produce arbitrary rotations in 
Hilbert space of any individual qubit, ie the propagator
  \beq
\exp( - i \mbox{\boldmath $\theta \cdot  \sigma$}/2 ) = 
\left( \begin{array}{rr}
\cos(\theta/2) & -e^{-i\phi} \sin(\theta/2) \\
e^{i \phi} \sin(\theta/2) & \cos( \theta/2 ) \end{array} \right)
  \eeq
and to be able to carry out the `controlled-rotation' operation {\sc crot} $= 
\ket{00}\bra{00} + \ket{01}\bra{01} + \ket{10}\bra{10} - \ket{11}\bra{11}$ 
between any pair of qubits. The notation used here is standard, the kets 
$\ket{0}$ and $\ket{1}$ refer to two orthogonal states of a qubit. This basis 
is referred to as the `computational basis', since this aids in designing 
useful algorithms for the QC. From a physical point of view, it is useful to 
take the computational basis to be the ground and excited eigenstates of the 
Hamiltonian of the relevent two-level system, though this is by no means 
required and any basis will serve. States
such as $\ket{01}$ are product states $\ket{0} = \ket{0} \otimes \ket{1}$ where 
the first written ket refers to one qubit, and the second another. For our 
purposes the qubits will always be distinguishable so we do not need to worry 
about the symetry of the states (with respect to exchange of particles) and any 
related quantum statistics. 

As mentioned previously, an operation like {\sc crot} is a propagator acting on 
the state of a pair of qubits. In matrix form it is written 
  \beq U_{\mbox{\sc crot}} = \left( \begin{array}{cccc}
1 \\ 
  & 1 \\
  &   & 1 \\
  &   &   & -1 \end{array} \right)
  \label{crot}  \eeq
in the basis $\ket{00}, \ket{01}, \ket{10}, \ket{11}$, where matrix
elements which are zero have not been written. The appellation `controlled
rotation' comes from the fact that if the first qubit is in the state 
$\ket{0}$, {\sc crot} has no effect, whereas if the first qubit is in the
state $\ket{1}$, {\sc crot} rotates the state of the second by the Pauli
$\sigma_z$ operator. 

The two operators just described form a universal set, which means that any 
possible unitary transformation can be carried out on a set of qubits by 
repeated use of these operators or `quantum gates', applied to different qubits 
\cite{ugate}. Another commonly considered quantum gate is the `controlled not' 
or `exclusive or' ({\sc xor}) gate
  \beq
U_{\mbox{\sc xor}} = \left( \begin{array}{cccc}
1 \\ 
  & 1 \\
  &   & 0 & 1 \\
  &   & 1 & 0 \end{array} \right),
\eeq
see also equation (\ref{xor}). This gate has no 
effect if the first qubit is in the state $\ket{0}$, but applies a {\sc not} 
operation ($\sigma_x$ Pauli spin operator) to the second qubit if the first is 
in the state $\ket{1}$. In the computational basis, this means that the state 
of the second qubit becomes the {\sc xor} of the two input qubit values. We 
have introduced {\sc crot} before {\sc xor} in this discussion, going against 
standard practice, because we shall see later that {\sc crot} is easier to 
implement in an ion trap. 

It should be emphasised that this model in terms of quantum gates operating
on quantum bits is by no means the only way to think about quantum computation,
but is the way which is most well understood at present, and is certainly
very powerful. Other models include those based on cellular automata,
and simulated annealing. 

A further simplification of the physical construction of a quantum computer is 
as follows. Instead of seeking a means to carry out {\sc crot} between any 
pair of qubits directly, it is sufficient to have one special qubit which can 
undergo {\sc crot} with any of the others. This special qubit acts as a 
one-bit `bus' to carry quantum information around the computer, 
making repeated use of the {\sc swap} operation $\ket{00}\bra{00} + 
\ket{10}\bra{01} + \ket{01}\bra{10} + \ket{11}\bra{11}$. To carry out {\sc 
crot} between any pair of qubits $x$ and $y$, one makes use of the bus bit $B$ 
as follows: $\mbox{\sc crot}(x,y) = \mbox{\sc swap}(B,x)\cdot\mbox{\sc 
crot}(B,y)\cdot\mbox{\sc swap}(B,x)$. The operation {\sc swap} can be built out 
of three {\sc xor}'s with the order of the bits alternating:
$\mbox{\sc swap}(B,x) = \mbox{\sc xor}(B,x)\cdot\mbox{\sc 
xor}(x,B)\cdot\mbox{\sc xor}(B,x)$, however in practice this construction is 
unnecessarily complicated, since {\sc swap} can be applied more or less 
directly in most physical implementations. 

The use of a bus bit makes the physical construction of a quantum 
information processor much simpler, and indeed most current proposals use 
this concept. However, it has the major disadvantage that more than one gate 
(acting on different sets of qubits) cannot be carried out simultaneously 
(ie in parallel), except single qubit rotations. Accepting this limitation, 
the minimum requirement for our processor is arbitrary rotations of any 
single qubit, plus {\sc crot} and {\sc swap} between the bus qubit and any 
of the others. This is the minimum set of `computing operations', in the 
sense that arbitrary transformations can be carried out by means of this 
small set. However, this establishes neither that arbitrary transformations 
can be carried out {\em efficiently}, nor that they can be carried out {\em 
without uncorrectable errors}, both of which are important additional 
considerations for a computer. We will return to these issues in the 
sections \ref{s:effic} and \ref{s:ecorr}. 

A further ingredient for quantum information processing is that the result of 
the process---here the final state of the quantum system---must be able to be 
measured without errors. A basis is chosen (typically the eigenbasis of the 
system Hamiltonian) and a measurement of all the qubits is carried out in this 
basis. 

To make a modest processor (a few qubits) the easiest approach is probably to 
use single particles with several internal degrees of freedom. Examples are a 
spin $J=2^{K-1}-1/2$ in a magnetic field (say $J=7/2$ giving $2J+1 = 8$ 
dimensions and therefore $K=3$ qubits); a molecule or confined particle with 
$2^K$ accessible vibrational states (`accessible' in this context means the 
experimenter can cause computing operations among the states at will). This 
approach will be interesting in the short term. However it is difficult to 
imagine it being extended in the longer term to enable the realisation of a 
really interesting processor with hundreds of qubits. Also, it is not clear how 
to apply arbitrary operations to a single particle (spin, molecule) with an 
evenly spaced ladder of energy levels, owing to level degeneracies in the
interaction picture.

There are now several proposed physical systems which might one day make a 
quantum computer \cite{Lloyd93,Berm94,DiV95,CZ,Pel95}. We will concentrate on 
the system of a line of ions in an ion trap, since it appears to be the most 
promising at present. However, developments in solid state physics may overtake 
us, and one should bear this in mind. It is not easy to couple the quantum 
information out of an ion trap system (ie in the form of qubits, not classical 
measurements), which is important for quantum communication. In this regard the 
approach based on strong coupling between an atom and a cavity mode appears 
more useful, since there a bit of quantum information could in principle be 
transferred into the polarisation state of a photon which then leaves the 
system in a chosen direction (a `flying qubit') \cite{Kimble}. However, such 
ideas could be applied to trapped ions, making a form of hybrid processor, so 
the ion trap system remains an interesting candidate even for quantum 
communication purposes. 

\section{Ion Trap Method}  \label{s:trap}

For reviews and references on ion trapping, see for example
\cite{Thomp93,revtrap,ps,Itan95,Ghosh}. The ion trap system which interests us 
uses a line of $N$ trapped ions. Each ion has two stable or metastable
states, for example two hyperfine components of the electronic ground state 
(which usually requires an odd isotope), or two Zeeman sublevels of the ground 
state, separated by applying a magnetic field. The ground state and a 
metastable electronic exited state (eg a D state for ions of alkaline earth 
elements) might also be used, but this is a poor choice since the laser 
linewidth and frequency, as well as most of the mirrors etc on the optical 
bench, will have to be very precisely controlled for such an approach to work. 
There have been optimistic estimates of the computational abilities of an ion 
trap processor, based on the use of such optical transitions, but one should 
beware of the lack of realism in such estimates. This will be discussed more 
fully once we have seen exactly how the system is intended to operate. 

There are $N$ laser beam pairs, each interacting with one of the ions,
(or a single beam which can be directed at will to any chosen ion), see
figure {1}. Each ion provides one qubit, the 
two-dimensional Hilbert space being spanned by two of the ion's
internal energy eigenstates.
A further $N+1$'th qubit acts as a `bus' enabling
the crucial {\sc crot} operations. This qubit is the vibrational motion
of the whole ion string in the trap potential. This motion must be quantised,
in other words the ion cloud temperature must be reduced well below the
`quantum limit' defined by the axial vibrational frequency in the ion trap:
  \beq
k_B T \ll \hbar \omega_z   \label{qlimit}
  \eeq
The first major experimental challenge (after making a trap and catching your 
ions) is to cool the ions down to this quantum regime. Note that the quantum
regime for the trapped motion of the ion is {\em not} related
to the ``Lamb-Dicke'' regime which will be considered below. In brief,
it will be shown that one wants to operate well into the quantum regime,
but on the border of the Lamb-Dicke regime.

So far the quantum regime has only been achieved for a single ion of either 
Mercury in two dimensions \cite{Died89} or Beryllium in three dimensions 
\cite{coolBe}. 
Both experiments used optical sideband cooling in the resolved sideband 
(tight trapping) limit. This and other possible cooling techniques will be 
discussed. 
Traps for neutral atoms have also attained the motional ground state, most 
spectacularly in the case of Bose Einstein condensation \cite{BEC}, but also in 
optical lattices \cite{latt}. These systems do not (at present) provide full 
control of individual atoms and interactions between pairs, so we will not 
discuss them. However, they lend further weight to the
impression that it is in atomic physics and quantum optics, rather than
solid state devices, that quantum information processing will be most
fruitful in the immediate future.

To get to the quantum regime, it appears to be neccessary to use a Paul
rather than Penning trap, since rf technology allows tighter confinement
than does high magnetic field technology. Therefore only the Paul trap
(rf trap) will be considered from now on, although we may permit ourselves
to add a magnetic field if we wish, for some other reason such as to enhance
the stability or split the Zeeman levels. In any case, tighter confinement
enables a faster `switching-time' for quantum gates such as {\sc crot},
so the tightest possible trap is the best option.

Note that once more than a single ion is in a three-dimensional rf trap of 
standard geometry (with the rf voltage between end caps and a ring) matters 
are complicated since no more than one ion can be at the centre of the trap 
potential. Away from the centre, ions undergo rf micromotion and this causes 
heating if there is more than one ion, due to collisions (Coulomb repulsion) 
which force the micromotion out of quadrature with the rf field. To avoid 
this, one must use a linear or ring geometry. The confinement along the axis 
is then either due to a static field from end cap electrodes (linear case), 
or to repulsion between ions combined with their confinement to a ring 
shape. In this case, only radial micromotion is present, but this vanishes 
for all the ions if they lie along the axis at the centre of the radial 
potential, so rf heating is avoided. The ring case must imply a small 
micromotion tangential to the ring, since the tangential and radial 
confinement can't be completely decoupled, but as far as I know this has not 
yet been found to be a problem. 

\subsection{Average motion}  \label{s:string}

We will model a row of $N$ ions in a trap as a system of $N$ point charges 
in a harmonic potential well of tight radial confinement, ie $\omega_x, 
\omega_y \gg \omega_z$, see figure {2}. The oscillation frequencies 
$\omega_x, \omega_y$ and $\omega_z$ are parameters which will be obtained 
from the electrode geometry and potentials in section \ref{s:rf}. The total 
Hamiltonian is 
  \beq
H = \sum_{i=1}^N \frac{1}{2} M \left( 
\omega_x^2 \hat{X}_i^2 +
\omega_y^2 \hat{Y}_i^2 +
\omega_z^2 \hat{Z}_i^2
+ \frac{{\bf \hat{P}}_i^2}{M^2} \right) 
+ \sum_{i=1}^N \sum_{j > i} \frac{e^2}{4 \pi \epsilon_0 \left|\hat{\bf R}_i
- \hat{\bf R}_j \right|}
  \label{Hline}  \eeq
For $\omega_x, \omega_y \gg \omega_z$ and at low temperatures, the ions all 
lie along the $z$-axis, so we can take $|\hat{\bf R}_i - \hat{\bf R}_j| 
\simeq |\hat{Z}_i - \hat{Z}_j|$ and the radial and axial motion can be 
separated. The axial motion interests us, so the problem is one-dimensional. 
A length scale is given by 
  \beq
z_s = \left( \frac{e^2}{4 \pi \epsilon_0 M \omega_z^2} \right)^{1/3}
  \label{z0}  \eeq
which is of the order of the separation between the ions
(typically 10 to 100 $\mu$m). Solving
the classical equations of motion (ie the operators $\hat{Z},\;\hat{P}_z$
become classical variables $z,\; p_z$) one obtains the
equilibrium positions shown in figure {3}. With more
than two trapped ions, the outer
ions tend to push the inner ones closer togther, so the
ion positions depend on $N$ (see equation (\ref{zmin})).
Remarkably, however, the
frequencies of the first two normal modes of oscillation about these
equilibrium positions are independent of $N$ (for
small oscillations) \cite{CZ}, and those of higher modes are
nearly independent of $N$. The frequencies of the first two modes
are $\omega_z$ and $\sqrt{3}\, \omega_z$, and those of higher modes
are given approximately by the list 
$\{1,\; \sqrt{3},\; \sqrt{29/5},\; 3.051,\; 3.671,\; 4.272,\;
4.864,\; 5.443,\; 6.013,\; 6.576\}$, which gives the frequency of
the highest mode, in units of $\omega_z$, for $N=1$ to $10$. 
The near independence of $N$ of the mode frequencies is illustrated
by figure {4}.

The lowest mode of oscillation corresponds to harmonic motion of
the centre of mass of the ion string. In this mode, all the ions
move to and fro together. It is important that the frequency of
this mode is significantly different from that of any other mode,
since this means that experimentally one can excite the centre
of mass mode without exciting any of the others.

We can now proceed directly to a quantum mechanical treatment, simply by 
treating the centre of mass coordinate $z_{\rm cm}$ as a harmonic oscillator. 
The classical result that the centre of mass normal mode has frequency 
$\omega_z$ remains valid even though the ion wavefunctions may now overlap, 
since all the internal interactions among the ions cancel when one calculates 
the centre of mass motion. Since we have an oscillator of mass $N M$ and 
frequency $\omega_z$, the energy eigenfunctions are
  \beq
\psi_n\left(z_{\rm cm}\right) =
\left( \frac{N M \omega_z}{\pi \hbar 2^{2n} (n!)^2} \right)^{1/4}
H_n\left( z_{\rm cm} \sqrt{N M \omega_z/\hbar}\, \right)
e^{-N M \omega_z z_{\rm cm}^2 / 2 \hbar} 
 \label{psi} \eeq
The spatial extent of the Gaussian ground state probability distribution
is indicated by its standard deviation
  \beq
\Delta z_{\rm cm} = \sqrt{\hbar / 2 N M \omega_z}.   \label{deltaz}
  \eeq
Since we wish a different laser beam to be able to address each of the ions, we
require $\Delta z_{\rm cm}$ to be small compared to
the separation between ions. The closest ions are those at the centre of
the line. A numerical solution of (\ref{Hline}) yields the following
formula for the separation of the central ions:
  \beq
\Delta z_{\rm min} \simeq 2.0 \, z_s N^{-0.57}.
  \label{zmin}  \eeq 
This formula is plotted for $N \le 10$ in figure {3}. An approximate
analytical treatment for $N \gg 1$ does not predict a power-law dependence of
$\Delta z_{\rm min}$ on $N$, but rather $\Delta z_{\rm min} \propto 
z_s (\log(N)/N^2)^{1/3}$ \cite{Dub93}. However, (\ref{zmin}) is more accurate
for $N < 10$ and remains accurate for the range of $N$ which interests
us (up to, say, $N=1000$). Setting $\Delta z_{\rm cm} \ll \Delta z_{\rm min}$
yields
  \beq
\frac{\omega_z}{M} \ll \frac{32 N^{1.86}}{\hbar^3}
\left( \frac{e^2}{4 \pi \epsilon_0} \right)^2
\simeq 2.4 \times 10^{21} \; N^{1.86} \; {\rm Hz}/{\rm u}    \label{overlap}
  \eeq
where u is the atomic mass unit $1.66057 \times 10^{-27}$ kg. This condition is 
easily fulfilled in practice, with $\omega_z$ no greater than a GHz, and $M$ 
between 9 and 200 u. Therefore it is legitimate to picture the ions as 
strung out in a line, each sitting in a small wavepacket centred at its 
classical equilibrium position, not overlapping the others. Note that
(\ref{overlap}) does not guarantee that the ions are sufficiently separated
to be addressed by different laser beams, only that their wavefunctions
do not overlap.

In the above it was assumed that the radial confinement was sufficient
to cause the ions to lie along the $z$-axis, rather than form a zigzag
or helix about it. The onset of such zigzag modes has been studied
numerically \cite{Schiff93} and analytically \cite{Dub93}. They occur when the 
ions approach sufficiently closely that the local potential minimum at the 
position of an ion on the $z$ axis becomes a saddle point. For a string
of ions uniformly spaced by $\Delta z$ (which is not the case in our harmonic 
trap), the transition from a line to a zigzag occurs when \cite{Tot88}
$\omega_r^2 \simeq 4.2072 (z_s / \Delta z)^3 \omega_z^2$, where
we have taken the case $\omega_x = \omega_y \equiv \omega_r$.
Setting $\Delta z = \Delta z_{\rm min}$, this leads to the condition
  \beq
\frac{\omega_r}{\omega_z} > 0.73 N^{0.86}   \label{zigzag}
  \eeq
for the prevention of zigzig modes. For $N \gg 1$, an approximate
analytic treatment yields the condition \cite{Dub93}
   \beq
\frac{\omega_r}{\omega_z} > 0.77 \frac{N}{\sqrt{ \log N }}.  \label{zigzag2}
   \eeq
These numerical and approximate analytic formulae are
within 10\% agreement for $3 < N < 2000$.

\subsection{Principle of operation}  \label{s:princ}

The principle of operation of an ion trap `information processor' was described 
by Cirac and Zoller \cite{CZ}, and the most important elements of such a system 
were first realised in the laboratory by Monroe {\em et al} \cite{Mon}. Whereas
the transition operators given by Cirac and Zoller were calculated for 
standing-wave excitation of allowed single-photon transitions, experimentally 
Monroe {\em et al} employed travelling-wave excitation of two-photon Raman 
transitions (cf figures {1} and {8}). The basic form of the operators is 
independent of the type of excitation used, however. The method may be 
understood by reference to figure {5}, which shows the relevant energy 
levels for one of the ions in the trap. We consider three of the ion's internal 
energy eigenstates $\ket{F_1,M_1},\;\ket{F_2,M_2}$ and $\ket{F_{\rm aux}, 
M_{\rm aux}}$, and various excitations of the centre of mass motion. The ion's 
internal energy levels are separated in frequency by $\omega_{0}$ and 
$\omega_{\rm aux}$ as indicated on figure {5}. Note that all these 
levels are low-lying, separated from the ground state only by hyperfine and 
Zeeman interactions (see figure {8}), so their natural lifetime against 
spontaneous emission of rf photons is essentially infinite. Figure {5} 
shows the lowest-lying excitations of the second, third and fourth normal modes 
as well as the first, to act as a reminder of the location of the closest 
extraneous levels whose excitation we wish to avoid. The energy eigenstates of 
the vibrational motion may be written $\ket{n_1, n_2, n_3, \ldots}$ where the 
$n_i$ are the excitations of the various normal modes. Only the ground state 
$\ket{0,0,0,\ldots}$ and first excited state of the centre of mass 
$\ket{1,0,0,\ldots}$ will be involved in the operations we wish to invoke. This 
centre of mass vibrational degree of freedom is often referred to somewhat 
loosely as a `phonon'. The `computational basis' consists of the states 
  \begin{eqnarray}
\ket{0,0} &\equiv & \ket{F_1,M_1} \otimes \ket{0,0,0,\ldots} \nonumber \\
\ket{0,1} &\equiv & \ket{F_1,M_1} \otimes \ket{1,0,0,\ldots} \nonumber \\
\ket{1,0} &\equiv & \ket{F_2,M_2} \otimes \ket{0,0,0,\ldots} \nonumber \\
\ket{1,1} &\equiv & \ket{F_2,M_2} \otimes \ket{1,0,0,\ldots}
\label{cbasis}
  \end{eqnarray}

It will now be shown
how to carry out {\sc crot} between any single ion's internal state
and the bus (phonon) bit, then how to carry out arbitrary rotations of
the internal state of an ion, then how to carry out {\sc swap} between
any ion and the bus bit. Recalling the discussion in section \ref{s:req}, 
these three operations form a universal set and so allow arbitrary
transformations of the stored qubits in the processor.

The auxillary states $\ket{{\rm aux},i} \equiv \ket{F_{\rm aux}, M_{\rm aux}} 
\otimes \ket{i,0,0,\ldots}$ ($i=0,1$) are available as a kind of `shelf'
by means of which useful state-selective transformations can be carried
out among the computational basis states. If one applies radiation at
the frequency $\omega_{\rm aux} + \omega_z$, then inspection of
figure {5} will reveal that only transitions between $\ket{1,1}$
and $\ket{{\rm aux},0}$ will take place (assuming that unwanted levels such as 
$\ket{1,2}$ are unoccupied)\footnote{Cirac and Zoller originally proposed to 
produce the selective effect of this {\sc crot} operation by means of a chosen 
laser polarisation rather than frequency. However, experimentally frequencies 
can be discriminated more precisely than polarisations, which explains why 
Monroe {\em et. al.} chose to use a frequency- rather than 
polarisation-selective method.}. 
If one applies a $2 \pi$ pulse at 
this frequency then the state $\ket{1,1}$ is rotated through $2 \pi$ radians, 
and therefore simply changes sign. In the computational basis, the effect is 
equal to that of the {\sc crot} operator described in section
\ref{s:req}, see equation (\ref{crot}).

A $2 \pi$ pulse at frequency $\omega_0 - \omega_{\rm 
aux} - \omega_z$ also produces a controlled rotation, only now the minus
sign appears on the second element down the diagnonal of the unitary matrix,
rather than the fourth, causing a sign change of the component $\ket{0,1}$
rather than $\ket{1,1}$. This case will be called {\sc c}$_{\neg}${\sc rot},
the negation symbol $\neg$ referring to the fact that
here the second qubit is rotated if the first is in the 
state $\ket{0}$ rather than $\ket{1}$.

To rotate an ion's internal state without affecting the centre of mass
motion, one applies radiation of frequency $\omega_0$. If such
radiation has phase $\phi$ with respect to some defined origin of phase,
and duration sufficient to make a $p\pi$ pulse, then the effect
in the computational basis is
  \beq
V^p (\phi) \equiv \left( \begin{array}{rr}
\cos(p\pi/2) & -i e^{-i \phi} \sin(p\pi/2) \\
-i e^{i \phi} \sin(p\pi/2) & \cos(p\pi/2) 
\end{array} \right)_{\rm ion} \otimes
\left( \begin{array}{cc} 1 & 0 \\ 0 & 1 \end{array} \right)_{\rm cm}
\label{Vphi}  \eeq
where we have followed the notation of \cite{CZ}, but used $p$ instead
of $k$ to avoid confusion with the wave vector.
Note that to apply such rotations succesfully, it is necessary to have
the phase of the radiation under experimental control. 
That means under control at the position of the ion, not just in some
stable reference cavity. This constitutes
a severe experimental constraint which makes computational basis
states separated by radio frequencies highly advantageous compared to states 
separated by optical frequencies. On the other hand, in order to have the right 
phase experimentally, note that one need not worry about the continuous 
precession at frequency $\omega_0$ caused by the internal Hamiltanian of each 
ion. The laser field keeps step with this precession, as becomes obvious when 
one uses the interaction picture, which we have done implicitly in writing 
equation (\ref{Vphi}). Problems arise when different ions have
different internal energies, due to residual electric and magnetic fields
in the apparatus, but such problems are surmountable.

The centre of mass motion acts as the `bus' qubit described in section 
\ref{s:req}. To carry out {\sc xor}$(B,x)$, between the `bus' and
the internal state of a single trapped ion,
Monroe {\em et. al.} applied first 
a $\pi/2$ pulse at frequency $\omega_0$, 
  \beq
V^{1/2}(-\pi/2) = \frac{1}{\sqrt{2}} \left( \begin{array}{rr}
 1 & 1 \\ 
-1 & 1 \end{array} \right)_{\rm ion} \otimes
\left( \begin{array}{cc} 1 & 0 \\ 0 & 1 \end{array} \right)_{\rm cm}
  \eeq
followed by {\sc crot} as described in the paragraph after equation 
(\ref{cbasis}), followed by a second $\pi/2$ pulse at $\omega_0$ with phase 
displaced by $\pi$ with respect to the first,
ie $V^{1/2}(\pi/2)$\footnote{In fact Monroe {\em et. al.} state that they
used $V^{1/2}(\pi/2)$ for the first pulse, and $V^{1/2}(-\pi/2)$ for the
third, producing {\sc xor} with an additional a rotation of the cm state.}. 
A straightforward calculation shows that this sequence produces exactly
  \beq
\mbox{\sc xor}({\rm cm},\; {\rm ion}) = \left( \begin{array}{cccc} 
1 & 0 & 0 & 0 \\
0 & 0 & 0 & 1 \\
0 & 0 & 1 & 0 \\
0 & 1 & 0 & 0   \end{array} \right)
  \label{xor}  \eeq

By symmetry, to obtain {\sc xor}$($ion, cm$)$, one might imagine using a 
similar sequence, but with
the $\pi/2$ pulses applied at frequency $\omega_z$ 
so as to affect the vibrational state without affecting the internal state. 
However, the vibrational degree of freedom is not really a two-level system, so 
this will not work (indeed, it will cause unwanted multiple excitations of the 
vibrational motion). To perform {\sc crot}, we made use of a transition at 
frequency $\omega_{\rm aux} + \omega_z$. Note that this relied on the fact that 
there was no population in the state $\ket{{\rm aux},1}$ (which would have 
become coupled to $\ket{1,2}$ which is outside the computational Hilbert 
space). This illustrates the general method by which the vibrational state is 
influenced: one uses radiation at a frequency offset from an internal resonance 
of the ion by $\omega_z$, thus coupling levels of vibrational quantum number 
differing by 1. To avoid coupling higher-lying vibrational states, one of the 
possible initial states must be unoccupied when such a transition is 
invoked. 

The transition at frequency $\omega_0 - \omega_z$ is indicated 
on figure {5}. A moment's reflection allows one to convince oneself 
that as long as there is no population in the $\ket{1,1}$ state (nor in 
extraneous states such as $\ket{0,2}$), application of this radiation will only 
cause transitions between $\ket{1,0}$ and $\ket{0,1}$, and hence a
{\sc swap} operation is available between the bus qubit and any other.
Applying a $p\pi$ pulse at phase $\phi$
and frequency $\omega_0 - \omega_z$, we obtain the operation
  \beq
U^p(\phi) \equiv \left( \begin{array}{rrrr}
1 &                         0 &                          0 & \times \\
0 &              \cos(p\pi/2) & -i e^{-i\phi} \sin(p\pi/2) & \times \\
0 & -i e^{i\phi} \sin(p\pi/2) &               \cos(p\pi/2) & \times \\
0 &                         0 &                          0 & \times
\end{array} \right)
  \eeq
where the final column of crosses indicates that an initial state $\ket{1,1}$
is carried out of the computational basis by $U^p(\phi)$. The case $p=1$,
that is a $\pi$ pulse, produces a {\sc swap} operation with an additional
$-i$ phase factor, which we will write $U^1(0) =$ {\sc swap}$_{(-i)}$. 
Applying $U^{1}(0)$ to ion $x$, followed by {\sc c$_{\neg}$rot} to ion $y$ 
(ie using the frequency $\omega_0 - \omega_{\rm aux} - \omega_z$), followed 
by $U^{1}(0)$ once again to ion $x$, has the effect of a {\sc crot} 
operation between $x$ and $y$. That is, {\sc crot}$(x,y) = 
\mbox{\sc swap}_{(-i)}(B,x) \cdot \mbox{\sc c$_{\neg}$rot}(B,y) \cdot
\mbox{\sc swap}_{(-i)}(B,x)$, as 
long as the initial state of $x$ and the bus is not $\ket{1,1}$. To apply the 
method, one uses the bus as a `work bit' which is arranged always to return to 
state $\ket{0}$ before operations such as $U^p(\phi)$ are applied, so the 
quantum information processing can go forward without problem.\footnote{Indeed, 
the bus may even be measured at those times when it should be in the ground 
state $\ket{0}$, producing a slight stabilisation or error detection, see 
section \protect\ref{s:ecorr}. In the ion trap, however, one can only thus 
measure the vibrational state by first swapping it with the internal state of a 
prepared ion.} 

So far we have described operations on the ion trap by means of $p\pi$ pulses. 
A complimentary technique is that of adiabatic passage, in which a quantum 
system is guided from one state to another by a strongly perturbing Hamiltonian 
applied slowly. For example, instead of swapping one ion's internal state with 
the bus qubit, and then swapping the bus with another ion, one could swap the 
internal state of two ions `via' the bus but without ever exciting the first 
vibrational level. The details are described for a related system in 
\cite{Pel95}. This method has experimental advantageous in being insensitive to 
features such as the timing and interaction strength. Both $p\pi$ pulses and 
adiabatic passage will probably have their uses in a practical QC. 

The laser pulses described provide the universal set of `quantum logic gates'
for the linear ion trap.
To complete the operation of our processor or QC, we require that the final 
state of the quantum information processor can be measured with high 
accuracy. This is possible for trapped ions by means of the `electron 
shelving' or `quantum jumps' technique \cite{Thomp93,ps,Ghosh}. That 
is, one may measure whether a given ion is in state $\ket{F_1,M_1}$ or 
$\ket{F_2,M_2}$ by illuminating it with radiation resonant with a transition 
from $\ket{F_1, M_1}$ to some high-lying level, whose linewidth is small 
enough so that transitions from $\ket{F_2,M_2}$ are not excited. If 
fluorescence is produced (which may be detected with high efficiency), the 
ion state has collapsed to $\ket{F_1, M_1}$, if none is produced, the ion 
state has collapsed to $\ket{F_2, M_2}$. 

\subsection{Efficient gate sequences} \label{s:effic}

It was shown in the previous section that {\sc crot} can be applied to any 
pair of qubits, and arbitrary rotations of single qubits can be carried out. 
Hence, as explained in section \ref{s:req}, any arbitrary sequence of 
unitary transformations of the quantum processor can be brought about. 
However, the most efficient methods will not blindly adopt a simple 
repetition of {\sc crot}'s and rotations to solve any problem. There may be 
much more efficient methods, by using other possible pulse sequences. Cirac 
and Zoller emphasize this by demonstrating how to apply a {\sc c}$^n${\sc 
rot} operation, in which the $\sigma_z$ operator is applied to one ion's 
internal state only if $n$ other ions are in the state $\ket{1}$, using a 
number of pulses equal to $2(n-2) + 3$. This is efficient in that the number 
rises only linearly with $n$, and the multiplying factor is small (ie 2 
rather than 48 as in \cite{Bar95}). 

Efficiency in computer science has a rigorous definition. Without going into 
details, the essential point is that if the number of elementary computational 
steps (here, quantum gates) required to complete an algorithm rises 
exponentially with the size of the input to the algorithm, then the algorithm 
is inefficient. The definitions can be made rigorous, which we will not attempt 
to do, but essentially each algorithm addresses not one instance of a problem, 
such as to ``find the square of 2357'', but a whole class of problems, such as, 
``given an integer $x$, find its square''. The `size of the input' to the 
algorithm is measured by the amount of information required to specify $x$, 
which is the number of digits in the binary expression of $x$, ie $\log_2(x)$. 
A computation is inefficient if the number of steps is exponential in 
$\log(x)$, ie is proportional to $x$. Similarly,
a quantum gate involving $n$ qubits is inefficient if the number of
physical operations, such as laser pulses, required to implement it is 
exponential in $n$ (eg increases as $2^n$). The strict definition of the 
universal computer mentioned in section \ref{s:req} also involves this 
efficiency aspect: when a universal computer simulates the action of 
another, the number of operations in the simulation algorithm must not rise 
exponentially with the amount of information required to define the 
simulated computer. 

Although we emphasised in section \ref{s:req} that a small set of
gates is `universal' in that all unitary transformations can be
composed by them, this does {\em not} necessarily imply that they can be used 
to build the particular transformations we may want in an efficient way. In 
this sense, the word `universal' is misleading. 

So far, networks of quantum gates have been designed for the most part 
without regard to the exact physical process which might underlie them. 
However, in such an approach it is not obvious which gates to call 
`elementary', since a physical system like the ion trap may be particularly 
amenable to some transformations. We have already seen an example in the 
{\sc swap} gate in the ion trap, which can be carried out without recourse 
to a sequence of {\sc xor} gates. This implies that a thorough understanding 
of a particular system like the ion trap may lead to progress in finding 
efficient networks. The important insight in Cirac and Zoller's construction 
of the {\sc c}$^n${\sc rot} gate is that the method makes use of $\pi$ 
pulses at frequency $\omega_{\rm aux} + \omega_z$. In other words, during 
the implementation of this gate the ions are deliberately carried out of the 
computational Hilbert space. Alternatively, one could regard the `shelf' 
level $\ket{F_{\rm aux}, M_{\rm aux}} \otimes \ket{i,0,0,\ldots}$ as within 
the computational Hilbert space, in which case we have more than one qubit 
available per ion. Later, in section \ref{s:ecorr}, we will consider
using vibrational modes in addition to the
lowest one in order to have more than one `bus' qubit.

\subsection{Switching Rate}  \label{s:rate}

The previous section showed how the ion trap information processor
worked, by invoking radiation of prescribed frequency and duration
in the form of $p \pi$ pulses. The `switching rate' of the
processor is limited by the duration of these pulses.

Let $\Omega$ be the Rabi frequency for resonant excitation of the internal
transition at frequency $\omega_0$ for a free ion. This will be
determined by the linestrength of the transition and the laser
power available. For a two-level atom one has $\Omega^2 = 6\pi\Gamma I/\hbar c 
k^3$ where $\Gamma$ is the linewidth of the transition, $I$ is the intensity
of the travelling wave exciting the transition, and $k$ is the
wavevector. When considering excitations of the internal state alone
of an ion in a trap, ie $\Delta n = 0$ where $n$ is the vibrational
quantum number, this `free ion' Rabi frequency still applies.
However, when changes in the vibrational state are involved,
ie transitions at frequency $\omega_0 \pm \omega_z$
producing $\Delta n = \pm 1$, an additional
scaling factor $\Delta z_{\rm cm} k_z$ appears, where $\Delta z_{\rm cm}$
is the extent of the ground state vibrational wave function given in
equation (\ref{deltaz}), and $k_z = k \cos(\theta)$ is the wavevector
component along the $z$ direction. Using (\ref{deltaz}), we have
  \beq
\Delta z_{\rm cm} k_z = \left( \frac{\hbar k^2 \cos^2(\theta)}
{2 N M \omega_z} \right)^{1/2} \equiv \frac{\eta}{\sqrt{N}}
  \eeq
where $\eta$ is the Lamb-Dicke parameter for a single trapped ion. In
the case of weak excitation, the 
effective Rabi frequency for the vibrational-state-changing transitions is 
$\eta \Omega / \sqrt{N}$, a result which can be interpreted as arising from 
conservation of momentum. The factor $\sqrt{N}$ appears because the whole ion 
string moves {\em en masse} and therefore has an effective mass $N M$ 
(M\"ossbauer effect). The Lamb-Dicke parameter can also be written
in terms of the recoil energy (energy of recoil of an ion after emission
of a single photon)
  \beq
E_R \equiv \frac{(\hbar k)^2}{2M},       \label{ER}
  \eeq
giving $\eta \equiv \cos(\theta) (E_R/\hbar \omega_z)^{1/2}$.

We can now obtain a measure of the switching rate $R$ by taking it as the 
inverse of the time to bring about a $2 \pi$ pulse on a 
vibrational-state-changing transition, ie 
  \beq
R \simeq \frac{\eta \Omega}{2 \pi \sqrt{N}}
\;\;\;\;\;\;\; (\eta \le \sqrt{N})          \label{tswitch}
  \eeq
Outside the Lamb-Dicke limit (ie for $\eta > \sqrt{N}$) the ion-radiation 
interaction is more or less equivalent to that of a free ion, so the factor 
$\eta/\sqrt{N}$ is replaced by 1.

It was remarked in the previous section that to maintain phase control between 
(and during) computing operations, there is a strong advantage in having the 
transition frequencies $\omega_0, \omega_{\rm aux}$ in
the rf to microwave rather 
than optical region of the electromagnetic spectrum. However, if the relevant 
transitions are driven directly by microwave radiation, with a frequency of the 
order of the vibrational frequency $\omega_z$, then the Lamb-Dicke parameter is 
extremely small (of order $(\hbar \omega_z/2 M c^2)^{1/2}$, so 
vibrational-state-changing transitions are almost impossible to drive. One way 
to avoid this would be to make the trap extremely weak, but this has the 
disadvantage of making the system sensitive to purturbations and lowering the 
switching rate. Instead, it is better to drive the microwave transitions by 
Raman scattering at optical frequencies. This combines the advantage of a large 
photon momentum and hence strong driving of vibrational-state-changing 
transitions, with the possibility of accurate phase control since only the 
phase {\em difference} between the pair of laser beams driving a Raman 
transition need be accurately controlled. The Raman technique was adopted for 
these reasons by Monroe {\em et. al.} \cite{Mon}. The same reasoning leads to 
the advantage of Raman scattering for precise laser-manipulation of free atoms 
\cite{Kas91,Reich}. A clear theoretical analysis is provided by \cite{Moler}. 

The maximum switching rate is dictated by the 
three frequencies $\Omega, \omega_z$ and $E_R/\hbar$ in a subtle way. If only 
low laser power is available, $\Omega \ll \omega_z$, then the Rabi frequency 
limits the switching rate and the best choice for $\omega_z$ is that which 
makes $\eta \sim \sqrt{N}$, ie 
   \beq
\eta^2 \equiv \frac{\cos^2(\theta) E_R}{\hbar \omega_z} \simeq N  \label{eta}
   \eeq
Therefore the recoil energy, given by the choice of 
ion and transition, dictates the choice of trap strength, for a given number of 
ions. Typical recoil energies for an ion are in the region $E_R \sim 
2\pi \hbar \times (10$--$200)$ kHz, and traps with this degree of confinement 
are now standard. In this situation, increasing the number of ions
does not affect the switching rate, but reduces the required trap 
confinement, making the system more sensitive to perturbations.

If higher Rabi frequencies are available, one's intuition suggests that 
$\omega_z$ becomes the limit on the switching rate, since $\Omega$ must be less 
than $\omega_z$ or the power broadening will no longer allow the different 
vibrational levels to be discriminated. However, at high $\omega_z$ one has 
$\eta \ll 1$ (Lamb-Dicke regime) so the switching rate on $\Delta n = \pm 1$ 
transitions cannot reach $\omega_z$ if $\Omega < \omega_z$. Placing the {\em ad 
hoc} limit $\Omega < \omega_z / 10$ in equation (\ref{tswitch}), one obtains 
  \beq
R < \frac{1}{20\pi} \left( \frac{E_R \omega_z}{\hbar N} \right)^{1/2}.
  \label{Rmax}  \eeq
The switching rate is thus limited by the geometric average of 
$\omega_z$ and $E_R / \hbar$, and the processor slows down when more ions are 
involved. For example, to achieve a switching at the recoil
frequency, ie $R = E_R / 2\pi\hbar$, with $N=10$ ions, (\ref{Rmax})
implies $\omega_z = 1000 E_R/ \hbar$ and $\Omega = 100 E_R / \hbar$.
To keep the ions in a straight line, equation (\ref{zigzag})
requires $\omega_r > 5300 E_R / \hbar$ which is very hard to achieve
experimentally.

There is another problem with increasing $\omega_z$ in order to
increase $R$. When $\omega_z$ is large, $\eta \ll \sqrt{N}$, so the
transitions which do not change the vibrational state ($\Delta n = 0$)
are much more strongly driven by the laser than those that do
($\Delta n = \pm 1$, equation (\ref{tswitch})). This increases the unwanted
off-resonant driving of $\Delta n = 0$ transitions when $\Delta n = \pm 1$
transitions are invoked to perform quantum gates between an ion and the
phonon `bus'. 
In principle it should be possible to run an ion trap processor at rates of 
order $\omega_z$ by relaxing the condition $\Omega < \omega_z$ and
allowing off-resonant transitions,
but the simple analysis given in section \ref{s:princ} is 
then no longer valid. One can no longer use a two-level model for each 
transition of the ion/centre-of-mass system. The a.c. Stark effect (light 
shift) will be all-important, and different computational basis states 
will become mixed by the ion--light interaction. The optical Bloch equations 
remain solvable (numerically if not analytically), and a detailed analysis 
should still enable useful elementary computing operations to be identified. 
Such an analysis is a possible avenue for future work. 

\section{Cooling}  \label{s:cool}

To make the quantum information processor described in the previous sections, 
the main initial requirements of an experimental system are cooling to the 
quantum regime, equation (\ref{qlimit}), and confinement to the border of the 
Lamb-Dicke regime, equation (\ref{eta}). The ions must be separated by at least 
several times the laser wavelength (equation (\ref{zmin})), but this is 
automatically the case, for small numbers of trapped ions, since with current 
technology the ions are always separated by many times the width of their 
vibrational ground state wavefunction (inequality (\ref{overlap})), which is 
itself approximately equal to the laser wavelength given that the Lamb-Dicke 
parameter is of order 1. 

Surveys of cooling methods in ion traps are given in \cite{Itan95,Ghosh}.
To cool to the quantum regime, there are two possible approaches.
Either one may cool to the ground state in the Lamb-Dicke
regime $\eta \ll 1$, then adiabatically open the trap to $\eta \sim 1$,
or one may apply cooling to a trap already at $\eta \sim 1$. The advantage
of the former approach is that one does not require cooling below
the recoil limit $k_B T_R = E_R$. The advantage of the latter is that
strong confinement is not necessary, but to attain the quantum
regime with $\eta \sim 1$ requires sub-recoil cooling. 

Cooling to the quantum regime has so far been demonstrated for trapped
ions by means of sideband cooling in the Lamb-Dicke limit
\cite{Died89,coolBe}.
This is described in section \ref{s:sideband} below. However, it
may be interesting to pursue other approaches, as discussed in
sections \ref{s:sis} and \ref{s:smcool}.

The physics of sideband cooling is very closely related to that
involved in information processing in the ion trap. This is no coincidence,
and a similar link will probably be found in all physical implementations
of quantum information processing. The relationship is sufficiently close
that one may say that once the goal of laser cooling to the motional
ground state is achieved in any given experimental ion trap, a
primitive form of quantum information processing can proceed
immediately, since all the required experimental components
will be in place. Conversely, 
quantum error correction (see section \ref{s:ecorr})
is a special type of `cooling'.

\subsection{Sideband Cooling}  \label{s:sideband}

Sideband cooling is just another name for the simplest type of
laser cooling, ie radiation pressure or Doppler cooling. The name
comes from how the process looks in the Lamb-Dicke limit $\eta \ll 1$.

There are several significant frequencies or energies. First, we have
the vibrational frequency in the ion trap potential, $\omega_z$. Next,
we have the radiative width of the transition used to do the cooling,
$\Gamma$. Either a single photon transition is used, in which case $\Gamma$ is 
its natural width (or possibly its broadened width if another laser is used to 
broaden a very narrow level as in \cite{Died89}), or a stimulated Raman
transition is 
used, in which case $\Gamma$ is some combination of the inverse of the duration 
of the Raman pulses, and the time for optical pumping out of one of the states 
linked by the Raman transition. Basically, the physics in the Raman case and 
single-photon case is very similar. The Raman method is a way of 
providing a very narrow transition when one is not already available. 
It also combines the advantages of precise frequency control (in the rf regime) 
with large photon recoil (optical regime), which permits fast cooling,
for the same reason that the switching rate for information processing
is faster (section \ref{s:rate}). One 
could instead use an rf or microwave transition, but then the cooling would be 
a lot slower and may not compete well enough with heating processes.

Laser cooling of atoms is often done quite happily using strong, resonant 
transitions. Indeed, such transitions are eagerly sought out. Why the talk of 
narrow transitions in the previous paragraph? It is because simple Doppler
cooling leads to the well-known Doppler cooling limit $k_B T_D \simeq
\hbar \Gamma / 2$, assuming the recoil energy is small compared to
$\hbar \Gamma$ (this applies in a trap as well as to free atoms). However,
we want to get to the quantum limit (equation (\ref{qlimit})), so
we require
  \beq
\omega_z \gg \Gamma.
\label{zG}  \eeq
This equation is a further constraint on the performance of the trap. It says 
the cooling transition must be narrow enough, or the trap confinement tight 
enough, to resolve the motional sidebands in the Lamb-Dicke spectrum. 

In the resolved sideband limit, radiation pressure cooling is called sideband 
cooling. A nice way of understanding it is to consider it as a form of optical 
pumping towards the state of lowest vibrational quantum number \cite{Itan95},
see figure {6}.
Note that the recoil after spontaneous emission produces heating. The average 
change in the vibrational energy per spontaneous emission is equal to the 
recoil energy $\hbar \omega_z \left< \Delta n \right> = E_R$ (a particularly
clear derivation of this fact may be found in \cite{CCT}). For a single
trapped ion illuminated by low-intensity light, the
cooling is governed by the following equation \cite{Wine79,CCT}:
  \beq
\frac{d}{dt}\left< H \right> = \frac{I \sigma_0}{\hbar \omega_L}
\sum_n P_n \sum_f \left(E_f - E_n + E_R \right)
\left| \left< \psi_n \right| e^{i \bf k \cdot R} \left| \psi_f \right>
\right|^2 g\left( \omega_L - (E_f - E_n)/\hbar \right)
 \label{cool} \eeq
where $I$ is the intensity of the incident radiation (a single travelling 
wave), $\sigma_0$ is the resonant photon scattering cross section ($\sigma_0 = 
2\pi \lambda^2 = (2\pi)^3 / k^2$ for a two-level atom), $\hbar \omega_L = \hbar 
c k$ is the laser photon energy, $P_n$ is the occupation probability of the 
$n$'th energy level of the vibrational motion, of energy $E_n = \hbar \omega_z 
(n+1/2)$ and wavefunction $\psi_n$ (equation (\ref{psi})), and $g(\omega)$ 
is the lineshape function. For a two-level atom,
  \beq
g(\omega) = \frac{\Gamma^2/4}
{\left( \omega - \omega_0 \right)^2 + \Gamma^2 / 4}
  \eeq

The quantity $d\left<H\right>/dt$ is the rate of change of the mean total 
energy of the ion, averaged over an absorption/spontaneous emission cycle. 
Since the ion's internal energy is left unchanged, this is the rate of change 
of the mean kinetic energy. Equation (\ref{cool}) has a simple physical 
interpretation as a sum of energy changes associated with
radiative transitions up and 
down the ladder of vibrational energy levels. At the lowest attainable
temperature, $d\left<H\right>/dt = 0$ and one possible solution of
equation (\ref{cool}) is the thermal distribution
  \beq
P_n = (1-s) s^n
  \eeq
where $s$ is the Boltzmann factor $s = \exp(-\hbar \omega_z / k_B T)$, and
the probability distribution has been properly normalised.

At sufficiently low temperatures, all but the lowest energy levels can be 
ignored in equation~(\ref{cool}). Using
${\bf k \cdot R} = \eta (\hat{a}^{\dagger} + \hat{a})$ 
where $\hat{a} \ket{\psi_n} = \sqrt{n} \ket{\psi_{n-1}}$,
and expanding in powers 
of the Lamb-Dicke parameter, it is a simple matter to obtain 
  \beq
\frac{d}{dt}\left< H \right> \simeq I \sigma_0 \frac{E_R}{\hbar \omega_L}
\left( \left< n \right> \left[ g( \omega_L - \omega_z )
- g( \omega_L + \omega_z ) \right] +
\left[ g(\omega_L) + g(\omega_L - \omega_z) \right] \right)
  \label{dHdt}  \eeq
where $\left<n\right> = \sum n P_n$ is the ion's mean vibrational quantum 
number. Now assume $\Gamma \ll \omega_z$ (inequality (\ref{zG})) and let the 
incident radiation be tuned to the first sideband below resonance, $\omega_L = 
\omega_0 - \omega_z$, then the cooling limit $d\left< H \right> / dt = 0$ leads 
to a mean vibrational quantum number~\cite{Neu78,Wine79}. 
  \beq
\left< n \right> \simeq \frac{5 \Gamma^2}{16 \omega_z^2}.
  \label{nmin}  \eeq
Note that since $\left< n \right>$ is proportional to $(\Gamma/\omega_z)^2$,
the experimental constraint (\ref{zG}) will ensure
achievement of the quantum limit $\left< n \right> \ll 1$.
This also justifies our ignoring higher energy levels
in deducing (\ref{nmin}).

The above assumed a single direction of propagation for the cooling laser, 
which will only result in cooling along one direction, so our calculation has 
been one-dimensional. Taking into account the fact that spontaneous photons 
are emitted into all directions, they do not heat any given dimension quite 
as much as we assumed, and the factor 5 in equation (\ref{nmin}) is replaced 
by $(1 + 4 \alpha)$ where $\alpha \simeq 2/5$ depends on the dipole radiation 
emission pattern \cite{Sten86}. However, this corresponds to an experiment in 
which the motion in the other dimensions is heated, which we wish to avoid. 
To cool all three dimensions, one can either introduce three laser beams, or 
use a single beam propagating at an oblique angle to all the principle axes 
of the trapping potential, and tune it separately to resonance with the three 
sideband frequencies $\omega_L - \omega_{x,y,z}$. For this one must have all 
three frequencies distinct, ie $\omega_x \neq \omega_y$. 

It is commonly imagined that sideband cooling is not possible if the recoil 
energy is greater than the phonon energy $\hbar \omega_z$, since then the 
cooling which results from photon absorption is undone by the recoil from 
photon emission, and $d\left< H \right> / dt > 0$. However, one can always 
tune to the next lower sideband, $\omega_L = \omega_0 - 2 \omega_z$, and 
good cooling is regained, as a thorough analysis of (\ref{cool}) will show. 
Therefore it is not necessary to be well into the Lamb-Dicke regime in order 
to attain the quantum limit by sideband cooling. 

Note also, that both equations (\ref{dHdt}) and (\ref{nmin}) are significant 
in order to find the minimum temperature one will obtain in the lab. This is 
because there will always be heating mechanisms present, such as a coupling 
between the stored ions and thermal voltages in the electrodes (see section 
\ref{s:perf}), so it is the cooling {\em rate}, equation (\ref{dHdt}), not 
just the minimum possible temperature, which is important. 

\subsection{Sisyphus cooling}  \label{s:sis}

The constraint (\ref{zG}) means that sideband cooling will either be slow and 
therefore not compete well with heating processes, or will require the use of 
Raman transitions. We can avoid $\Gamma \ll \omega_z$ and nevertheless use 
laser cooling to get close to the quantum regime, by the use of `Sisyphus' 
cooling \cite{Wine92,Cir92}. This makes use of optical pumping and optical 
dipole forces (forces associated with a position-dependent a.c. Stark shift 
of the atomic energy levels) in a laser standing wave, on an atom with at 
least three internal states. When the dipole force is caused by a 
position-dependent polarisation of the standing wave, the cooling is 
referred to as `polarisation gradient cooling'. Theoretical analyses 
\cite{Wine92,Cir92,Cir93} have so far concluded that the lowest temperatures 
attainable by this method correspond to a mean vibrational quantum number 
$\left< n \right> \simeq 1$, ie just on the border of the regime we require. 
However, the cooling rate is important as well as the theoretical minimum
temperature, 
and for this reason Sisyphus cooling may be attractive for cooling a whole 
string of ions \cite{Walt}, as required for the information processor, since 
it is relatively fast. A final stage of sideband cooling or something 
similar would then be required to get well into the quantum regime. 

\subsection{Statistical mechanical cooling methods} \label{s:smcool}

So far, all the cooling techniques described have been based on laser cooling. 
However, for trapped neutral atoms the technique of forced evaporative cooling 
has been shown to be extremely powerful, enabling the temperature in a weakly 
interacting atomic vapour to be brought well into the quantum
regime of a trap, which for a cloud of Bosons leads to Bose
Einstein condensation \cite{BEC}. 

In forced evaporative cooling, one starts with a large number of trapped 
particles in thermal equilibrium. Those of higher energy are forced to leave 
the trap, and those remaining rethermalise towards a lower equilibrium 
temperature. The technique relies on an ability to remove selectively particles 
of higher than average energy. One way to do this is to reduce the depth of the 
trap, allowing the faster particles to fly out. Clearly this approach will only 
work if the thermal energy is located more in some particles than in others, 
which is true for a gas of weakly interacting particles, but not for a 
crystalised system such as a cold string of trapped ions. However, evaporation 
may be useful in an ion trap as a first stage of cooling, to bring about 
crystalisation. Also, it is conceivable that a Bose condensate of neutral atoms 
may one day be sufficiently easy to produce in the vicinity of an ion trap that 
it may be used as a cold reservoir to cool the ions through collisions. The
use of one species to cool another is referred to as `sympathetic cooling'.

\section{rf requirements}  \label{s:rf}

We now turn to the design of the ion trap itself. The electrode
structure of the trap consists of a two-dimensional rf quadrupole plus
an axial static potential. Concentrating on the two-dimensional
quadrupole, consider first the most simple case, in which the
point in the centre of the electrode structure remains at zero potential,
and we omit any axial confinement.
The potential on one pair of diagonally opposed electrodes is
$(U - V \cos \Omega_V t)/2$, and that on the other pair has equal
magnitude and opposite sign to this. Here $\Omega_V$ is the frequency of 
the applied voltage, the subscript is necessary to distinguish it from the Rabi 
frequency of a driven atomic transition introduced in previous sections. The 
potential as a function of position in the $x$-$y$ plane is $\phi(x,y,t) = (U-V 
\cos \Omega_V t)(x^2 - y^2) / 2 r_0^2$ where $r_0$ is a measure of the 
electrode separation\footnote{In practice it is advantageous to
avoid exact cylindrical symetry in order to have all three vibrational
frequencies distinct, but this will unnecessarily complicate the
present discussion.}
For the case of cylindrical electrodes, $r_0$ is
the distance from the axis to the surface of the electrodes \cite{Raiz92}.
The trapping effect in the radial direction is 
stable as long as $\Omega_V$ is not too small, and is strong as long as 
$\Omega_V$ is not too large. This may be parametrised in terms of the standard 
parameters 
  \begin{eqnarray}
a &=& \frac{4 e U}{M r_0^2 \Omega_V^2}, \\
q &=& \frac{2 e V}{M r_0^2 \Omega_V^2},  \label{q}
  \end{eqnarray}
where $e$ is the charge on a trapped ion.
For present purposes, a zero dc potential difference $U=0$ may be used, so 
$a=0$. The radial confinement is then stable as long as $q$ is less than about 
$0.9$ \cite{Thomp93,Ghosh}. The radial micromotion has a velocity amplitude 
of $q \Omega_V \rho/2$ for an ion at average distance $\rho$ from the $z$-axis. 
The average motion on a time scale slow compared to $1/\Omega_V$, the so-called 
secular motion, can be modelled in terms of the pseudopotential
$\frac{1}{2} M \omega_r^2(x^2 + y^2)/e$ with radial vibrational frequency 
  \beq
\omega_r = \sqrt{a^2 + q^2/2} \; \frac{\Omega_V}{2} 
= \frac{q \Omega_V}{2\sqrt{2}} \;\;\;\;\;\;\;\;(a=0).
  \label{omegar}  \eeq
Choosing $q = 1/\sqrt{2}$ so as to be comfortably in the zone of stability
of the trap, we obtain $\omega_r = \Omega_V/4$. From this the Lamb-Dicke
parameter for the radial confinemt is obtained as
  \beq
\eta_r = \left( \frac{ 2 \sqrt{2}\; E_R k^2 r_0^2 }{e V} \right)^{1/4}
  \label{etar}  \eeq
where $k$ is the wavevector and $E_R$ the recoil energy as defined in
equation (\ref{ER}), and we have neglected the $\cos (\theta)$ term
for simplicity. The significance of equation (\ref{etar}) is that,
for a given ion and wavevector,
the Lamb-Dicke parameter of the radial confinement is dictated primarily
by the choice of electrode size ($r_0$) and rf voltage amplitude $V$. The 
required rf frequency $\Omega_V$ is dictated by $V/r_0^2$ through equation 
(\ref{q}) and the stability condition $q \simeq 1/\sqrt{2}$. 

Note that for information processing, we wish the Lamb-Dicke parameter
for the {\em axial} motion to be around 1, assuming there is only a small
number of ions in the trap. We also wish the ions to adopt the shape
of a linear string, so the radial confinement must be tighter than the axial
confinement (equation (\ref{zigzag},\ref{zigzag2})).
Taken together these two considerations 
imply that the Lamb-Dicke parameter for the {\em radial} motion should be 
much less than 1. 

Let us now add to the linear trap an axial dc potential, so that the ions 
are confined in all three dimensions, and with no axial micromotion. The 
most obvious way to do this is to add positively charged electrodes to 
either end of the linear trap, but this introduces a difficulty in correctly 
balancing the rf potential so that there is no residual axial rf component. 
An ingenious way around this is to split the linear electrodes of the radial 
quadrupole field and impose a potential difference between their two ends, 
as described in \cite{Raiz92}, see figure {1}. In either case, the 
dc potential near the centre of the trap will take the form of a harmonic 
saddle point potential 
  \beq
\phi_{\rm dc}(x,y,z) = \frac{U_z}{z_0^2}
\left[ z^2 - \mbox{\small $\frac{1}{2}$} \left(x^2 + y^2\right)\right]
  \label{phidc}  \eeq
where $U_z$ is the potential on each electrode, and $z_0$ is a parameter
which is measure of the electrode separation (its exact value depending
on the geometry). From this equation we obtain the vibrational frequency
for the axial harmonic motion of a trapped ion:
  \beq
\omega_z = \sqrt{ \frac{2 e U_z}{M z_0^2} }    \label{omz}
  \eeq
This is also the frequency of the lowest mode of vibration of a string of 
trapped ions (centre of mass mode), as discussed in section \ref{s:string}.
The Lamb-Dicke parameter of the axial confinement is
  \beq
\eta_z = \left( \frac{E_R k^2 z_0^2}{4 e U_z} \right)^{1/4}.
  \label{etaz}  \eeq

Owing to Earnshaw's theorem, it is impossible to apply an axial dc potential 
without influencing the radial confinement. The dc potential
$\phi_{\rm dc}(x,y,z)$ has the effect of expelling the ions in the radial 
direction. In the presence of both static axial and fluctuating radial electric 
potentials, the secular (ie slow) radial motion is still harmonic, but the 
vibrational frequency is no longer $\omega_r$ but 
  \beq
\omega'_r = \left( \omega_r^2 -
\mbox{\small $\frac{1}{2}$}\omega_z^2 \right)^{1/2}
  \eeq
However, as long as $\omega_r \gg \omega_z$, which is the case we are 
interested in, then $\omega'_r \simeq \omega_r$ so the previous discussion
of the radial confinement remains approximately valid, and in particular
the stability condition $q < \sim 0.9$ is not greatly changed. 

The depth of the trap (and hence the ease of catching ions) is given
approximately by the smaller of $e U_z$ and $e V/11$.

\section{Candidate Ions}   \label{s:ions}

Table 1 gives a list of ions which are suitable for information
processing. The list consists of ions whose electronic structure is
sufficiently simple to allow laser cooling without the need for too many
different laser frequencies. The list is not intended to be exhaustive,
but contains most ions which have been laser cooled in the laboratory. 

For information processing, a large recoil energy is attractive from the point 
of view of allowing a faster switching rate (equation (\ref{Rmax})), but makes 
the Lamb-Dicke regime harder to achieve (equations (\ref{etar}), (\ref{etaz})). 
The choice of rf rather than optical transitions for information processing
appears so advantageous as to be forced upon us. Since we
require at least three long-lived low-lying internal states of the ion
(the states $\ket{0},\;\ket{1}$ and $\ket{\rm aux}$), this implies that
the existence of hyperfine structure (ie a non-zero nuclear spin isotope), 
while complicating the cooling process, may be advantageous. Indeed, for 
alkali-like ions, (such as singly charged ions from group 2 of the
periodic table) a 
non-zero nuclear spin is required, since for zero nuclear spin
the total angular momentum of the ground state is only $1/2$, 
yielding only two long-lived states (the Zeeman components $\ket{J,M} = 
\ket{1/2,\pm 1/2}$) which is not sufficient. Most even isotopes have
zero nuclear spin.

The other major consideration is the difficulty in generating the light 
required for cooling and information processing. 

Examining figure {7} and table 1 it is seen that $^9$Be is an 
attractive choice, in that it allows the fastest switching rate, requires only 
one laser wavelength for cooling, and the hyperfine splitting frequency of 
$1.25$ GHz is accesible to electroptic modulators. However, the wavelength of 
$313$~nm requires the use of a dye laser (frequency doubled) which is 
disadvantageous. The next most promising candidate appears to be $^{43}$Ca. It 
requires two laser wavelengths for cooling, 397 (or 393)~nm and 866 (or 850)~nm 
(figure {8}), but both can be produced by diode lasers (one frequency 
doubled) which makes this ion very attractive (strontium has
similar advantages). Diode lasers can be made very 
stable in both frequency and power. If more laser power is needed than is 
possible with diode lasers, then a titanium-sapphire laser can be used, which 
is also advantageous compared with dye lasers. The hyperfine splitting of 
$3.26$~GHz is accessible to electrooptic modulators, though less easily than 
the smaller splitting in beryllium. The obvious difficulty in working with 
$^{43}$Ca is that it is a rare isotope, having a natural abundance of only 
$0.14$\% or 1 part in 700, making an isotopically enriched sample that much 
more expensive. However, one could carry out preliminary experiments using the 
$97$\% abundant $^{40}$Ca, in order to bring the trapping and cooling 
techniques up to performance, and the {\sc swap} operation could be tested 
since it does not make use of the auxilliary state $\ket{F_{\rm aux},M_{\rm 
aux}}$. 

For a group 2 ion, the internal states required for information processing, 
discussed in section \ref{s:princ} and illustrated in figure {5}, 
will be taken from the ground state hyperfine manifold. For $^{43}$Ca,
for example, one might take $\ket{F_1=4, M_1=4}, \;\ket{F_2=3, M_2=3}$
and $\ket{F_{\rm aux}=4, M_{\rm aux}=2}$. This choice is highlighted
in figure {8}. The degeneracy between 
the first and auxilliary levels is lifted by an imposed magnetic field
of order 0.1 mT.

\subsection{Example: the $^{43}$Ca$^+$ ion}  \label{s:eg}

To estimate laser power requirements, we will calculate the intensity required 
to saturate the $4 S_{1/2}$--$4 P_{3/2}$ transition in Ca$^+$ (for laser 
cooling purposes) and that required for Raman transitions in the ground state 
via a quasi-resonance with this transition (for information processing 
purposes). Using a two-level model for the allowed electric dipole transition, 
the saturation intensity (defined as the intensity giving a Rabi frequency 
equal to the FWHM linewidth $\Gamma$ divided by $\sqrt{2}\,$) is $I_{S} = 4 
\pi^2 \hbar c \Gamma / 6 \lambda^3 = 48$ mW/cm$^2$
(using $\Gamma = 2\pi \times 23$ MHz, $\lambda = 397$ nm).
To initiate laser cooling,
this intensity must be available in a laser beam wide enough to intersect a 
significant proportion of a `hot' ion's trajectory in the trap. Taking a beam 
diameter of 1 mm, the required laser power is of order 0.5 mW, which is a
large overestimate in practice.

Raman transitions from $\ket{0}$ to $\ket{1}$ via a near-resonance
with an excited state $\ket{e}$ can be modelled as transitions
in an effective two-level system, in which the effective Rabi frequency
of the Raman transition is 
  \beq
\Omega_{\rm eff} = \frac{\Omega_0 \Omega_1}{2 \Delta},  \label{Rabieff}
  \eeq
where $\Omega_0$ and $\Omega_1$ are the Rabi frequencies
of the single-photon transitions from levels $\ket{0}$ and $\ket{1}$
to $\ket{e}$, and $\Delta \gg \Omega_0, \Omega_1$ is the detuning from 
resonance of both of these transitions. Assuming level $\ket{e}$
only decays to levels $\ket{0}$ and $\ket{1}$, the single photon
Rabi frequencies can be obtained from the laser intensity $I$
and the linewidth $\Gamma$ of the excited
state, leading to $\Omega_{\rm eff} \simeq I \Gamma^2 / 8 I_S \Delta$.
During a Raman transition, the average population of the excited
state $\ket{e}$ is $\sim \Omega_0^2/4 \Delta^2$, and to produce,
for example, a $2\pi$ pulse, the pulse duration is $2\pi/\Omega_{\rm eff}$. 
Therefore the probability of an unwanted spontaneous emmision process
duration such an operation is 
  \beq
p_{\rm em} = \frac{\pi \Gamma}{\Delta}.  \label{pem}
  \eeq
An interesting possibility, which has not yet been tried in an ion trap,
is to use the Argon ion laser line at 488 nm to drive Raman transitions.
In this case, we have $\Delta \simeq 6 \times 10^6 \Gamma$, so
$p_{\rm em} \simeq 5 \times 10^{-7}$, allowing a million computing
operations before
spontaneous emission is a problem. The laser intensity required to obtain
$\hbar \Omega_{\rm eff} = E_R$ is then
  \beq 
I = 8 I_S \frac{\Delta}{\Gamma} \frac{E_R}{\hbar \Gamma}
\simeq 2.9 \times 10^9 \;\mbox{W/m}^2.
  \eeq
To address a single ion, the laser is focussed to a tight spot 
of diameter of order $10\;\mu$m, so the required power is modest, of
order $0.3$ W. 

Finally, to confine Ca to the Lamb-Dicke regime of the 393 nm radiation,
we require $\omega_z = E_R/\hbar \simeq 2 \pi \times 29$ kHz.
This is a reasonable choice for information processing, bearing in mind
the remarks made in section \ref{s:rate} about off-resonant transitions.
Choosing
an axial electrode separation of $\sim 4$ mm, the voltage required on the 
axial electrodes is $U_z \sim 0.12$ volts. This surprisingly low value 
arises from the fact that we have assumed the ions can be cooled to the 
ground state of the axial motion, which here corresponds to a temperature 
small compared to the recoil limit $E_R / k_B \simeq 1.4\;\mu$K. It shows 
that contact potentials will certainly be a problem, and one must be able to 
compensate them by seperately controlling the voltage on each electrode. To 
make the radial confinement 10 times stronger than this axial confinement we 
require an alternating voltage on the radial quadrupole electrodes of 
frequency $\Omega_V \simeq 2\pi \times 1.2$ MHz (equation (\ref{omegar})) 
and amplitude $V \simeq 9$ volts (equation (\ref{q})), assuming a distance 
of $\sim 1$ mm from the axis to the radial electrode surfaces.

\section{Performance limitations}  \label{s:perf}

Having begun in section \ref{s:princ} with an idealised treatment, in which we 
assumed operations could be carried out in an ion trap with arbitrary 
precision, the discussion has become in section \ref{s:ions} more realistic. It 
now remains to discuss the limitations on the performance of the ion trap 
system for information processing purposes.

Two important figures of merit for a quantum information processor are
the number of stored qubits, which so far in this paper has been the number
of trapped ions $N$, and the number $Q$ of elementary operations which can
be carried out before dissipation or decoherence causes a significant
loss of quantum information. To first approximation, we may
quantify dissipation or decoherence by a simple rate $\Gamma_d$, in which
case $Q = R / \Gamma_d$, where the switching rate $R$ is given in section
\ref{s:rate}. If we model decoherence as if each ion were independently
coupled to a thermal reservoir, leading
to a phase decoherence rate $\gamma$ for 
any individual ion, then we must take $\Gamma_d = N \gamma$ since the 
quantum computation is likely to produce entangled states in which the 
off-diagonal elements of the density matrix decay at this enhanced rate 
\cite{Car93,Cal85,Unruh95,Palma}. However, decoherence of a many-ion state 
in an ion trap is not yet sufficiently well understood to tell whether such 
a model applies \cite{Wine94}. Two possible thermal reservoirs affecting
the ion trap are electrical resistance in the electrodes, and thermal
radiation.

The major problems in an ion trap are spontaneous transitions in the 
vibrational motion, ie heating (a random walk up and down the ladder of 
vibrational energy levels), thermal radiation (driving internal rf
transitions in the ions),
and experimental instabilities such as in the laser 
beam power, rf voltages and mechanical vibrations, and fluctuating external 
magnetic fields \cite{Mon}. The instabilities contribute to the heating, and 
also imply that a laser pulse is never of exactly the right frequency and 
duration to produce the intended quantum gate. A `decoherence rate' figure of a 
few kHz was quoted by Monroe {\em et. al.} \cite{Mon}, consistent with the 
heating rate of 1000 vibrational quanta per second quoted in their earlier work 
\cite{coolBe}. With the switching rate of order 20 kHz, they obtained
$Q \simeq 10$ with $N=1$. A heating rate of 6 quanta per second was 
reported by Diedrich {\em et. al.} \cite{Died89}.

A useful model of the motion of a trapped ion is a series LC circuit which
is shunted by the capacitance of the trap electrodes \cite{Wine75,Wine94}. 
The inductance in the model is given by $l \simeq M z_0^2 / N e^2$
where $z_0$ is of the order of the axial electrode separation. A
resistance $r$ is due to losses in the electrodes and other conductors
in the circuit. This resistance both damps and heats the ionic motion
with time constant $l/r$, leading to a heating rate in vibrational
quanta per second \cite{Wine94}:
  \beq
\Gamma_{\rm heat} \simeq \frac{r}{l}\frac{k_B T}{\hbar \omega_z}
\simeq \frac{r N e^2 k_B T }{M z_0^2 \hbar \omega_z} 
  \label{heat}  \eeq
For example, substituting the parameters from section \ref{s:eg}, and using 
$r = 0.1$~ohm, $T = 300$~K, we obtain
$\Gamma_{\rm heat} = 5$ s$^{-1}$. It should be born in mind that one can
only consider equation (\ref{heat}) to apply once other sources of
electrical noise, such as rf pickup, have been reduced sufficiently, so
one cannot hope to improve the performance merely by increasing the
electrode separation $z_0$ and voltage $U_z$. 

It is a simple matter to combine equations (\ref{tswitch}), (\ref{omz}),
(\ref{Rabieff}) and (\ref{heat}) in order to obtain
$Q = R/\Gamma_{\rm heat}$ as a function of the experimental parameters.
However, this does not bring much insight and it is better to think
in terms of the switching rate and decoherence rate. Taking $N=10$
ions with $\omega_z = E_R / \hbar = 2 \pi \times 29$ kHz, as suggested
in the previous section, the switching rate is about 1 kHz, and $Q \sim 200$
using the value just quoted for $\Gamma_{\rm heat}$.
These parameters indicate what will probably be achievable in the next
few years. 

It is not hard to show that the influence of spontaneous emission of photons 
by the ions in the trap is much less important than the severe experimental 
problems just mentioned. Spontaneous emission takes place during the 
application of a laser pulse, due to the unavoidable weak excitation of an 
excited state of the relevant ion, as noted in the previous section. (It was 
already remarked that spontaneous emission between laser pulses is 
negligible, owing to the adoption of the ground state hyperfine manifold for 
computing). 
Taking the probability $p_{\rm em}$ from equation (\ref{pem}),
the number of operations that can be carried out before spontaneous
emmission plays a significant role is $Q \simeq 1/p_{\rm em}$ which can
be of the order of $10^6$, as remarked after equation (\ref{pem}).

The conclusion is that for the moment the limitations of the ion
trap are associated with the vibrational degrees of freedom, and
with experimental instabilities. It is here that experimental and
theoretical work must concentrate if progress is to be made. It remains
misleading at present to talk of quantum `computations' taking place
in the lab.

\subsection{Error correction} \label{s:ecorr}

Although it is important to build an information processor with as much 
precision and stability as possible, in the longer term the aim of significant 
computations is almost certainly unrealisable without something which goes 
beyond such `passive' stabilisation. It was initially thought that anything 
like active stabilisation of a quantum computer would be impossible, since it 
would rely on a means of monitoring the quantum state of the computer, which 
would irreversibly destroy the computation. However, the union of information 
theory with quantum mechanics has lead to another powerful concept, that of 
quantum error correction \cite{Cald,Steane2,Knill}. The essential idea is that 
$K$ qubits of quantum information in the quantum computer can be stored 
(`encoded') in a carefully chosen way among $N > K$ two-state systems. The 
computation is carried out in this specially chosen $2^K$-dimensional subspace 
of the total Hilbert space ($2^N$ dimensions) of the enlarged computer. The 
important feature is that the encoding is chosen so that the most likely 
errors, for example caused by heating of the vibrational degrees of freedom of 
the ion trap, cause the computer's state to go out of the special subspace. 
Such departures can be detected by well-chosen measurements on the computer, 
without upsetting the evolution of the quantum computation. Furthermore, a good 
encoding enables the most likely errors to be corrected, once they
are detected, by the application of one or more extra quantum gates
or dissipative measurements. 

Quantum error correction is more powerful than the more simple `watchdog 
effect' idea which preceded it \cite{zeno}.
The `watchdog' or `quantum Zeno effect' 
relies on rapid repeated measurements of the state of a system in order to 
prevent unwanted changes caused by the influence of some `error' Hamiltonian
$H_e$. It is successful if many measurements
can be applied within a time $t$ sufficiently small that
$1 - \left|\left<\phi\right| \exp(-i H_e t/\hbar )
\left| \phi \right>\right|^2   \propto t^2$.
In the midst of a computation one does 
not necessarily know what quantum state $\left|\phi\right>$ any
part of a QC should be in, so the watchdog method cannot be applied
directly, and a more subtle approach is required \cite{Bert94}.
However, it may not be
possible to make measurements rapidly enough, and it is not
clear whether the method can be applied during the application of a 
quantum gate, in which the quantum system is required to go through a 
prescribed evolution. 

By contrast with the Zeno effect, quantum error
correction allows a finite error term in the system's density 
matrix to accumulate, and corrects it afterwards. This is particularly 
important to the operation of quantum gates. For example, a gate between two 
qubits involves a four-dimensional logical Hilbert space. To allow error 
correction, we must ensure that at no point is the whole action of this gate 
concentrated into a four-dimensional physical Hilbert space. This can be done 
as follows. Suppose each qubit in the QC is encoded into two physical 
two-state systems. A gate $U(a,b)$ between two such encoded qubits $a,b$ can 
then be applied in four steps $U(a,b) = u(a_2,b_1) \cdot  u(a_1,b_2) \cdot 
u(a_2,b_2) \cdot u(a_1,b_1)$, where the operators $u$ are gates between a 
pair of two-state systems, and $\{a_1, a_2\}, \{b_1, b_2\}$ are the sets of 
two-state systems storing qubits $a$ and $b$. The important point is that 
error correction can be applied {\em between} the four $u$ operations.
At no stage is any quantum information stored in a physical
Hilbert space only just large enough to hold it, neither is any gate $U$
carried out in a single step. The proper combination of these features 
so as to allow stabilisation has
been dubbed `fault tolerance' \cite{Shor96}.

Quantum error correction was initially discussed with a general model 
of error processes in the quantum computer, in order to show that almost any 
imaginable error process might in principle be corrected by such techniques 
\cite{Shor95,Steane1,Cald,Steane2}. Subsequently, a technique 
specifically adapted to the vibrational noise in an ion trap was proposed 
\cite{Cir96}. In this proposal, the Hilbert space is enlarged by
making use of four internal states in each ion to store each qubit
in the quantum computation. This enables a two-qubit gate to be carried
out in four steps as outlined above. Next, we require a method of detecting 
and correcting the most likely errors in the vibrational state. This is done 
by using the first ($n=0$) and fourth ($n=3$) vibrational states, 
$\ket{0,0,0,\ldots}$ and $\ket{3,0,0,\ldots}$, instead of the first two, to 
store the `bus' qubit. The vibrational quantum number $n$ is measured 
whenever it should be $0$, by swapping the phonon state with the state of 
additional ions introduced for the purpose, and probing them. If $n$ is 
found to be $1$, then corrective measures are applied based on the 
assumption that a single jump upwards from $n=0$ occured, the details are 
given in \cite{Cir96}. If $n$ is found to be $2$ or $4$, then corrective 
measures are applied based on the assumption that a single jump down or up 
from $n=3$ occured. If $n$ is found to be $0$ as it should be, a corrective 
measure is still required to allow for the difference between such 
conditional evolution and the unitary evolution without jumps. This 
procedure enables single jumps up or down the ladder of vibrational levels 
to be corrected. Since these will be the most likely errors (at a sufficient 
degree of isolation from the environment), the effect overall is to 
stabilise the QC. In this case the figure of merit $Q$ is roughly squared 
(where $Q$ counts the possible number of logical gates $U$, not subgates 
$u$), a remarkable enhancement.

The above procedure makes allowance for the fact that errors cause the 
vibrational state to explore a Hilbert space of more than two dimensions, so 
in the language of quantum information, the bus size is larger than a single 
qubit, though the bus is still only used to store a single logical qubit. 
The bus could be made larger still by using higher excitations of the 
fundamental normal mode, or by using higher-order normal modes. This should 
allow more powerful error correction, and hence further increases in $Q$. 
The basic theory of error correction gives hope that 
such increases in $Q$ can be dramatic \cite{Cald,Steane2,Shor96}. 

Error correction should not be regarded as a device merely of interest to 
quantum computers. Rather, it is a powerful method of enforcing coherent 
evolution on a quantum system which would otherwise be dissipative. Such a 
capability may be useful for quite general situations in which stability 
is important, for example in low-noise electronic circuits, and frequency 
and mass standards. This may prove to be an area in which quantum 
information theory has provided a useful tool for other branches of 
physics. 

\section{Conclusion}

In conclusion, let us summarise the main avenues for future work
involving the ion trap quantum information processor. 

One of the basic aims of quantum information theory is to link abstract 
ideas on the nature information with the laws of physics. The ion trap 
provides a means of establishing this link in a complete and concrete way. 
This will set the theory on a more firm basis. 

An important task in the theory of quantum computation is that of 
identifying efficient multiple-qubit quantum gates. So far, it has been 
assumed that the efficient gates are those that can be divided into a set 
of sufficiently few two-qubit gates. However, a system like the ion trap 
may allow particular unitary transformtions to be carried out efficiently 
without dividing them into many two-qubit operations. An investigation
of this should be fruitful.

The principle of operation of the ion trap which we have described made 
use of various approximations whose influence on long quantum computations 
has yet to be analysed. For example, any given laser pulse on an ion will 
involve off-resonant stimulation of transitions other than the specific 
transition the pulse is designed to drive. Such effects may be unwanted, 
but their influence is unitary and accurately predictable. It would be 
interesting to investigate whether these effects can be taken into account 
in designing the sequence of laser pulses, so that they do not need to be 
corrected, or whether we are forced to regard them as errors. 

Quantum computation will certainly require error correction if it is ever 
to be useful for computational purposes. The ion trap provides a guide to 
the specific type of error correction which is likely to be required in 
the future. The basic tools of error correction are now fairly well 
understood, but there is much work to be done in bringing them to bear on 
the ion trap. In additition, these ideas may offer significant advantages 
for other uses of the ion trap, such as frequency and mass standards; this 
should be explored. 

Error correction only works once the level of noise in the trap is brought 
sufficiently low by careful construction and isolation. Experimental ion 
trap systems must be made much more stable than they are at present before 
they can take advantage of error correction of multiple errors among many 
qubits. This is not just a question of technology, but also of a better 
understanding of the noise processes, especially the influence of 
electrical noise in the electrodes providing axial confinement. The most 
immediate experimental challenge is to cool a many-ion crystal to the 
motional ground state. 

I would like to acknowledge helpful conversations with R. Thompson, D. Segal, 
D. Stacey and especially J. Brochard. The author is supported by the Royal 
Society. 
  
\newpage

\begin{figure} \label{f:piccy}
\end{figure}
Figure 1: Experimental arrangement. A line of three ions sits between
cylindrical electrodes, here seen sideways on. Pairs of laser beams
excite Raman transitions, which impart momentum changes to the ions
along the axial direction of the trap. 
The double-ended arrow indicates the direction of the resulting
oscillations, it can be regarded as a pictorial representation
of the fourth `qubit' in the system (see text, section
\protect\ref{s:princ}). The electrodes are split
in order to allow a constant voltage to be applied between their
ends, so that an axial potential minimum occurs
in the region where the long electrode segments overlap. Radial
confinement is provided by alternating voltages, see text (section
\protect\ref{s:rf}).

\begin{figure} \label{f:qpot}
\end{figure}

Figure 2: Schematic illustration showing an anisotropic harmonic
potential with the positions of three trapped ions indicated.

\begin{figure} \label{f:pos}
\end{figure}
Figure 3: Equilibrium positions for a line of point charges in a
quadratic potential, as a function of the number of charges $N$.
The positions are in units of $z_0$ (equation (\protect\ref{z0})).
The curve is equation (\protect\ref{zmin}).

\begin{figure} \label{f:modes}
\end{figure}
Figure 4: Normal mode frequencies for a line of point charges in
a quadratic potential, as a function of the number of charges $N$.
The frequencies are in units of $\omega_z$.

\begin{figure} \label{f:lev}
\end{figure} 
Figure 5: Energy levels and transitions in a single ion significant for 
information processing with a line of trapped ions. The labels $F,M$ 
indicate different internal states of the ion. Each internal state has an 
associated set of vibrational levels for each of the vibrational modes. 
Here, just the first few levels of the lowest mode (spacing $\omega_z$),
the first 
pair of levels of the next mode (spacing $\protect\sqrt{3} \,\omega_z$) and 
the lowest levels of two further modes are shown. The full arrows indicate 
transitions at frequencies $\omega_0-\omega_z$ and $\omega_{\rm 
aux}+\omega_z$, which are used in the {\sc swap}$_{(-i)}$ and {\sc crot} 
operations described in the text. Note that radiation at a given frequency 
couples not only the levels at the two ends of the relevant arrow on the 
diagram, but also other pairs of levels with the same difference of
vibrational quantum number. The figure shows $\omega_{\rm aux}$ to be of 
the same order as $\omega_0$, because this is what typically occurs when 
alkali-like ions are used. The vibrational frequency is smaller, $\omega_0 
\protect\simeq 100 \omega_z$. 
 
\begin{figure}  \label{f:cool}
\end{figure}
Figure 6: Sideband cooling. A laser excites transitions in which the 
vibrational quantum number of a confined ion falls by 1 (or a higher 
integer). Spontaneous transitions bring the ion's internal state back to the 
ground state, with the vibrational quantum number changing by $\pm 1$ or 0. 
On average the vibrational quantum number is reduced, until the vibrational
ground state is reached. The internal ground and excited states are
$\ket{g}$ and $\ket{e}$, and the figure shows the different vibrational
levels spread out horizontally for clarity. When both $\ket{g}$ and
$\ket{e}$ are long-lived (for example they may be the computational
basis states, cf figures 5 and 8), the $\ket{g,n} \rightarrow \ket{e,n-1}$
transition is driven by a $\pi$-pulse, ($U^1(0)$ operator), and the
spontaneous transition is a Raman transition via an unstable excited state
(optical pumping). Note that such experimental techniques are identical
to those required for information processing and error correction.
To cool a crystal of several ions, it is sufficient 
for the laser to interact with only one ion since the Coulomb coupling 
between ions causes rapid thermalisation of their motional state. However, 
the coupling between different normal modes is weaker, so these may need to 
be cooled separately by tuning the laser to the various normal mode 
sideband frequencies.

\begin{figure}  \label{f:ions}
\end{figure}
Figure 7: The recoil energies and main (typically S--P) transition wavelengths 
for ions which may be amenable to quantum information processing
(cf table 1).
A high recoil energy is advantageous for a high switching rate, 
but tends to be associated with a short wavelength. A rough rule is
that the shorter the wavelength, the more complicated and therefore less
stable is the laser system. The starred symbols are singly-ionised 
ions from group 1, in which a metastable manifold is used for
computing, making them unattractive in the long term.

\begin{figure}  \label{f:Ca}
\end{figure}
Figure 8: Low-lying energy levels of the $^{43}$Ca$^+$ ion, of
electronic structure $1{\rm s}^2 2{\rm s}^2 2{\rm p}^6 3{\rm s}^2
3{\rm p}^6 n l$. The hyperfine
structure is shown (nuclear spin $I = 7/2$), and the Zeeman sublevels and 
transition wavelengths. The levels are labelled by the value of the total 
angular momentum $F$, and a possible choice of computational and auxilliary 
levels is shown by the thickened Zeeman sublevels
in the ground state manifold. An 
example Raman transition is shown, for use both in sideband cooling and for the 
operation of quantum gates. The dashed transitions are forbidden by the 
electric dipole radiation selection rules in LS coupling, and so are weak. The 
3d$D$ levels are an unwanted complication; they must be depopulated by optical 
pumping during laser cooling. The level separations and lifetimes are taken 
from references \protect\cite{Caref}.

\begin{table} \end{table}
Table 1: List of candidate ions for information processing. 
Only singly-charged ions are considered, although some
ions of higher charge may also be interesting. For each
element, only the most abundant isotope, and those having
non-zero nuclear spin are shown. Unstable isotopes are not shown,
although most elements in the list (all but Mg, In and Li) have
further isotopes of half-life longer than one week.
The hyperfine splittings are for
the ground state in all but the inert-gas-like ions (Li, Na); they are
taken from G. Werth in \protect\cite{ps}.
The S--D wavelength is only shown when the D level lies below the P
level. The recoil energy is based on the S--P wavelength.
For Li, Na the S,P,D labels do
not apply; the transitions are from the metastable triplet state.

\newpage

\begin{table}
\begin{tabular}{|r|cccccc|}
\hline
Element, & Natural   & Nuclear & Hyperfine & $\lambda$
                                                & $\lambda$ & Recoil \\
isotopes & abundance & spin    & splitting & S--P  & S--D   & energy \\
         & (\%)    & $(\hbar)$ & (GHz)     & (nm)  & (nm)   & (kHz) \\
\hline
{\bf Be}   9 &   100 & 3/2 & 1.25001767 & 313 &     & 226 \\
\hline
{\bf Mg}  24 &    79 &   0 &            & 280 &     & 106 \\
          25 &    10 & 5/2 & 1.7887631  &     &     & \\
\hline
{\bf Ca}  40 &    97 &   0 &            & 397 & 730 & 30 \\
          43 &  0.14 & 7/2 & 3.25560829 &     &     & \\
\hline   
          87 &     7 & 9/2 & 5.00236835 &     &     & \\
{\bf Sr}  88 &    83 &   0 &            & 422 & 674 & 12.7 \\
\hline
         135 &   6.6 & 3/2 & 7.18334024 &     &     & \\
         137 &    11 & 3/2 & 8.03774167 &     &     & \\
{\bf Ba} 138 &    72 &   0 &            & 493 & 1760 & 5.94 \\
\hline
         199 &    17 & 1/2 & 40.507348  &     &     & \\
         201 &    13 & 3/2 &  ?         &     &     & \\
{\bf Hg} 202 &    30 &   0 &            & 194 & 282 & 26.6 \\
\hline
         113 &     4 & 9/2 & ?          &     &     & \\
{\bf In} 115 &    96 & 9/2 & ?          & 231 & 236 & 32.5 \\
\hline
         203 &    30 & 1/2 &   ?       &     &     & \\
{\bf Tl} 205 &   70  & 1/2 &   ?       & 191 & 202 & 26.8 \\
\hline
         171 &    14 & 1/2 & 12.6428121 &     &     & \\
         173 &    16 & 5/2 & 10.4917202 &     &     & \\
{\bf Yb} 174 &    32 &   0 &            & 369 & 411 & 8.42 \\
\hline
           6 &   7.5 &   1 & ?          &     &     & \\
{\bf Li}$^*$ 7 & 92.5& 3/2 & ?          & 539 &     & 94.7 \\
\hline                             
{\bf Na}$^*$ 23 & 100 & 3/2 & ?         & 249 &     & 140 \\
\hline
\end{tabular}
\end{table}
                                            
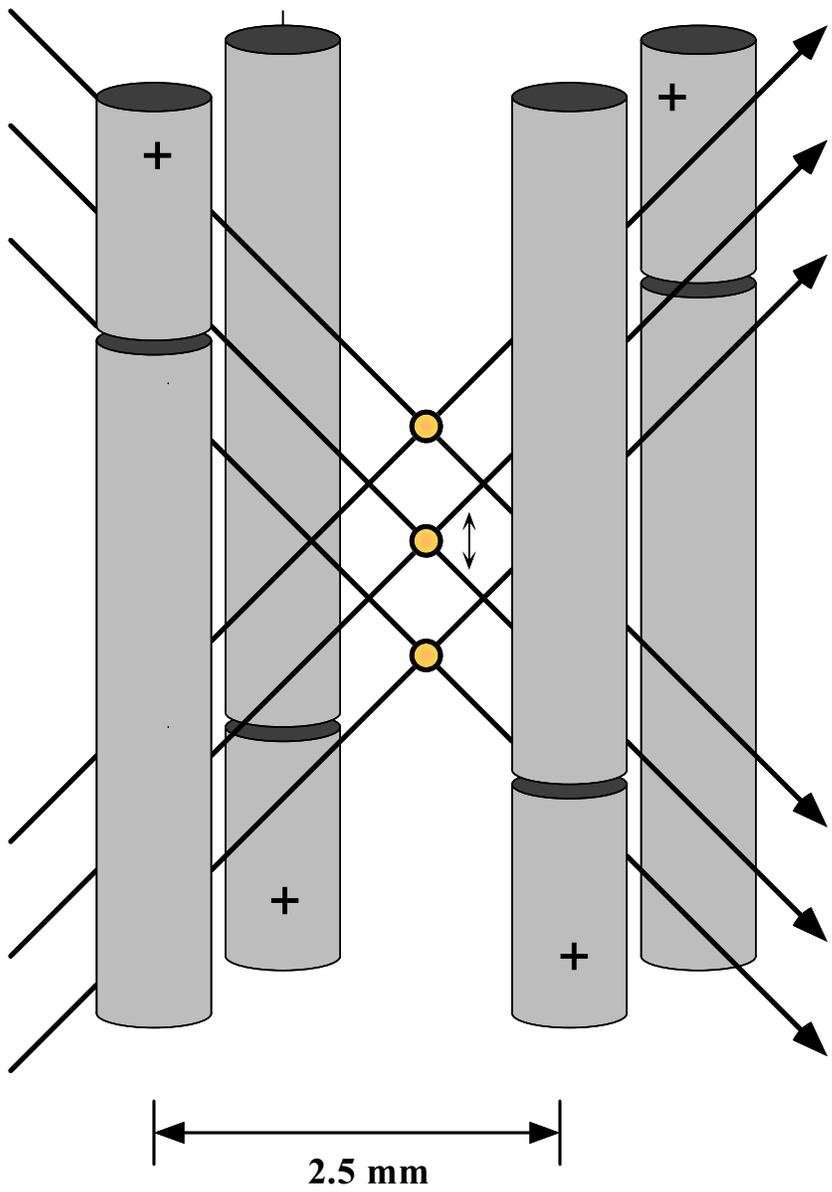
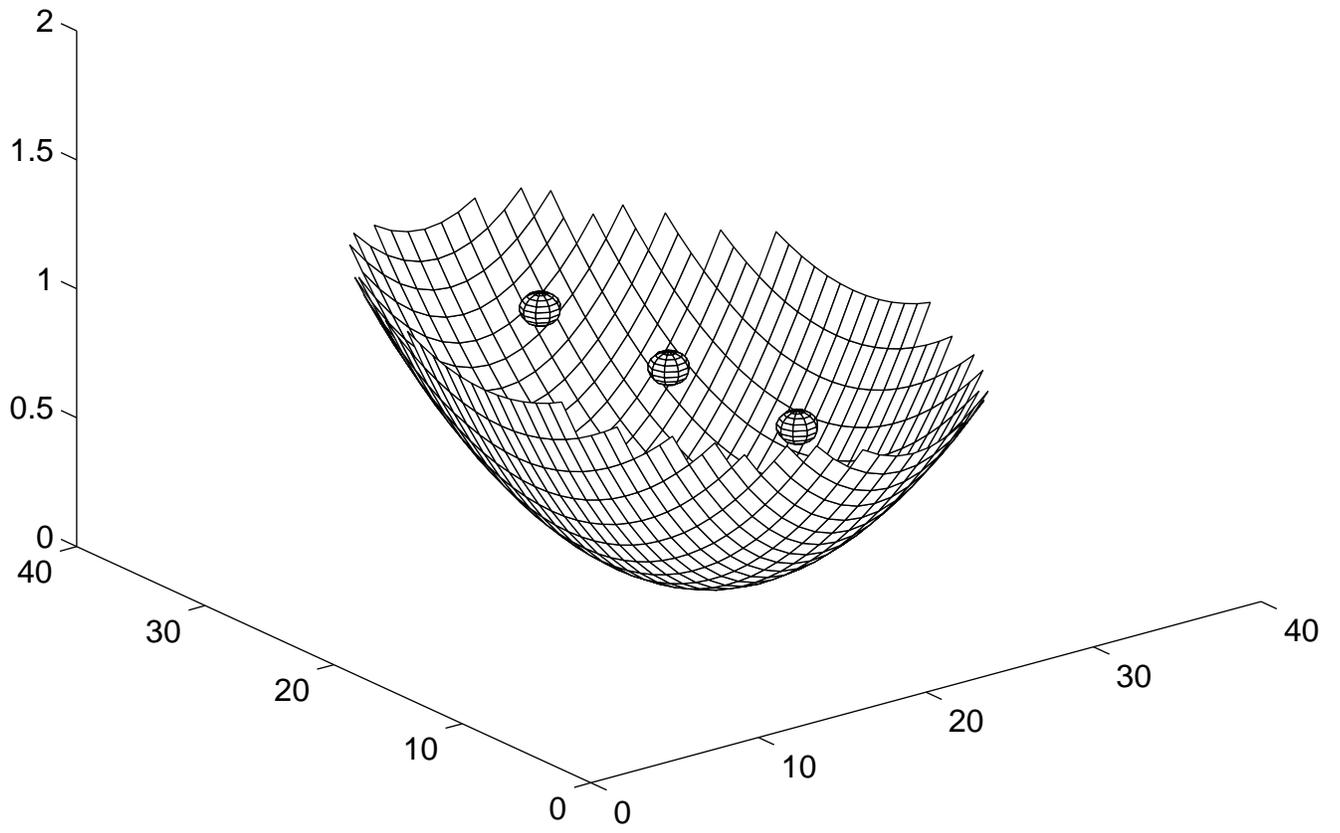
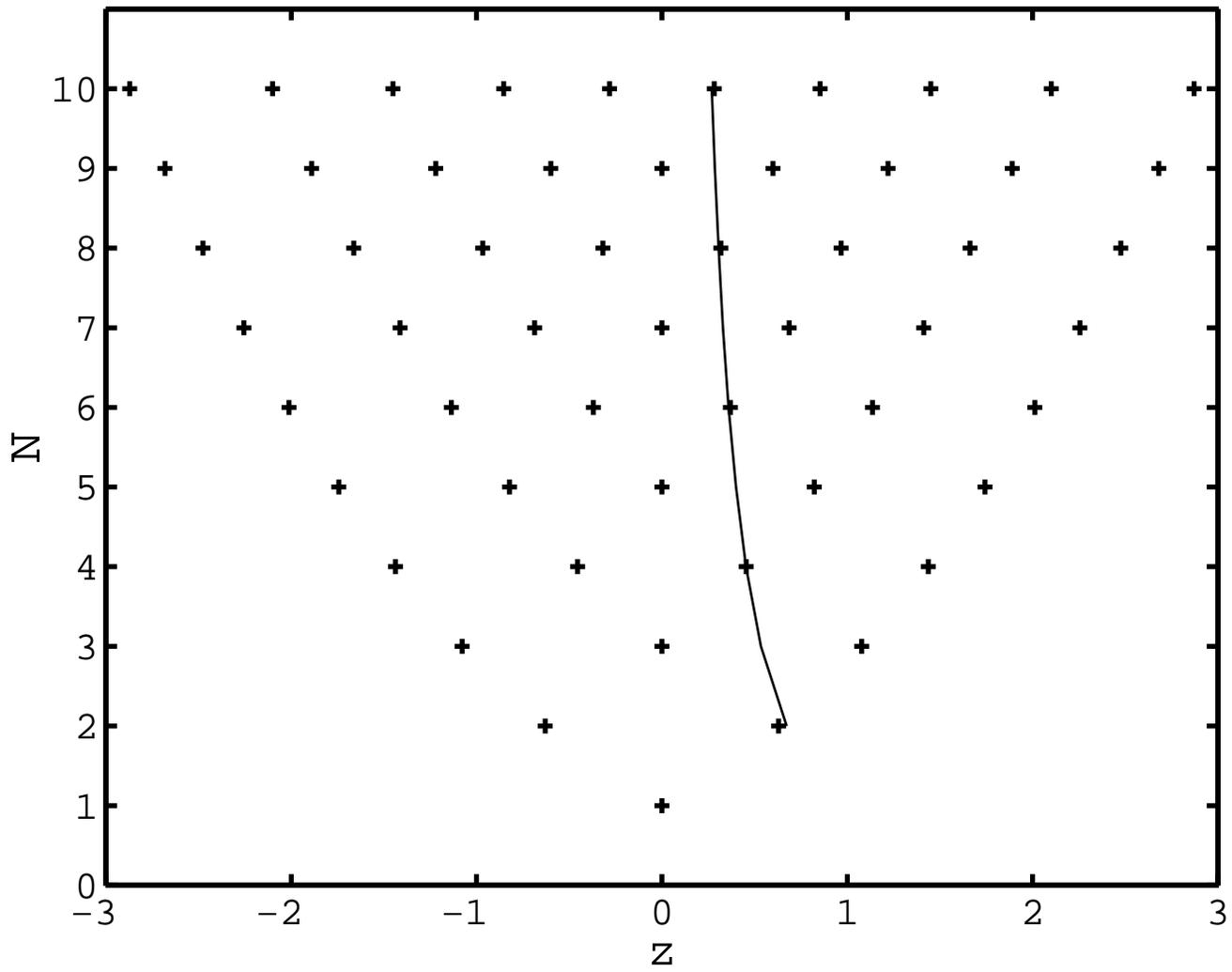
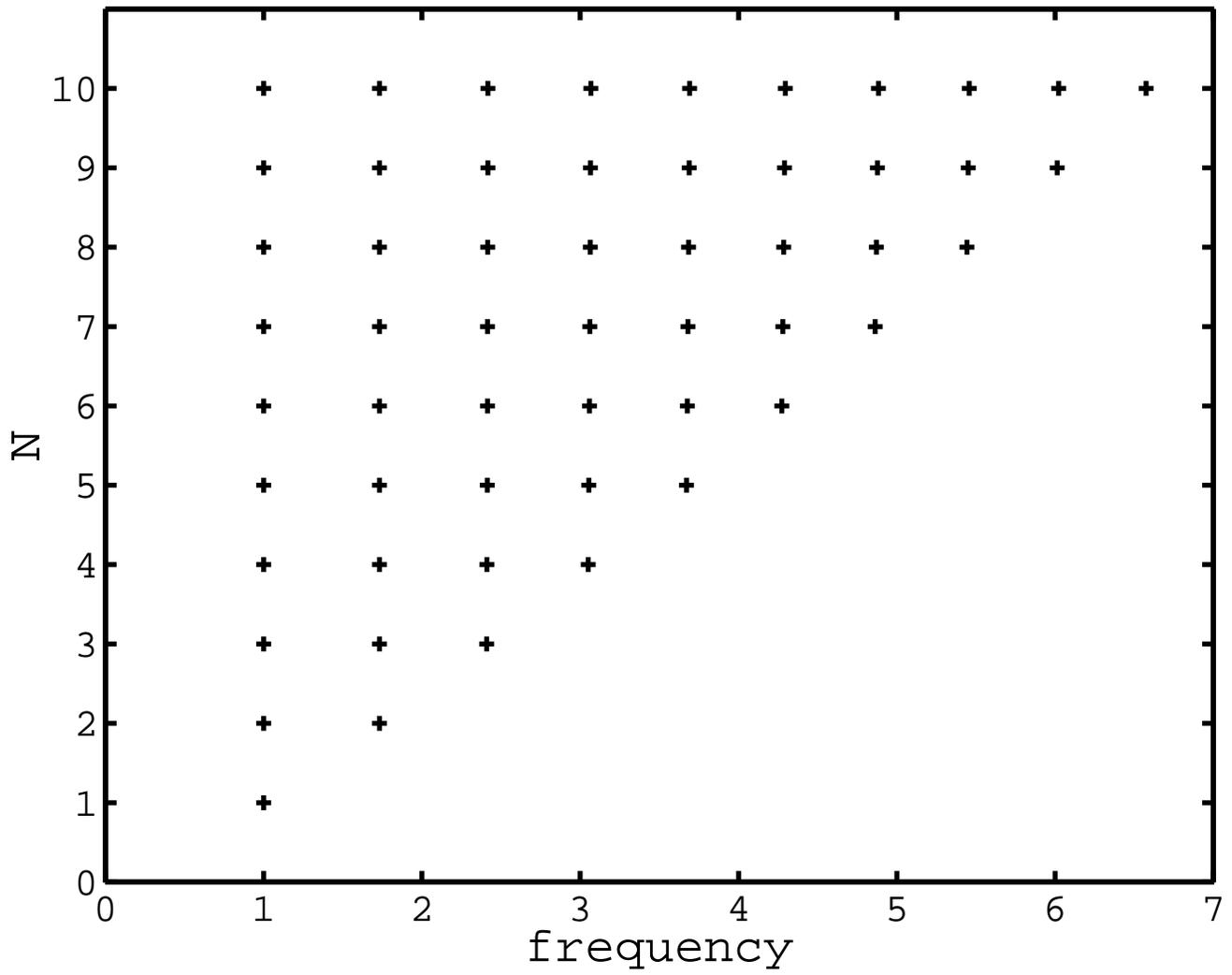
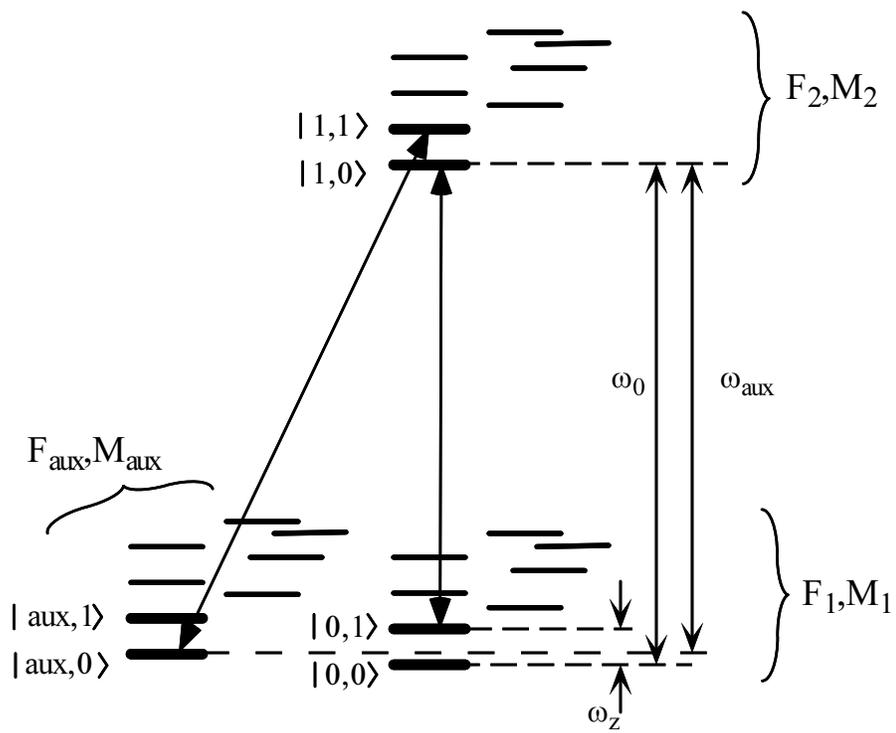
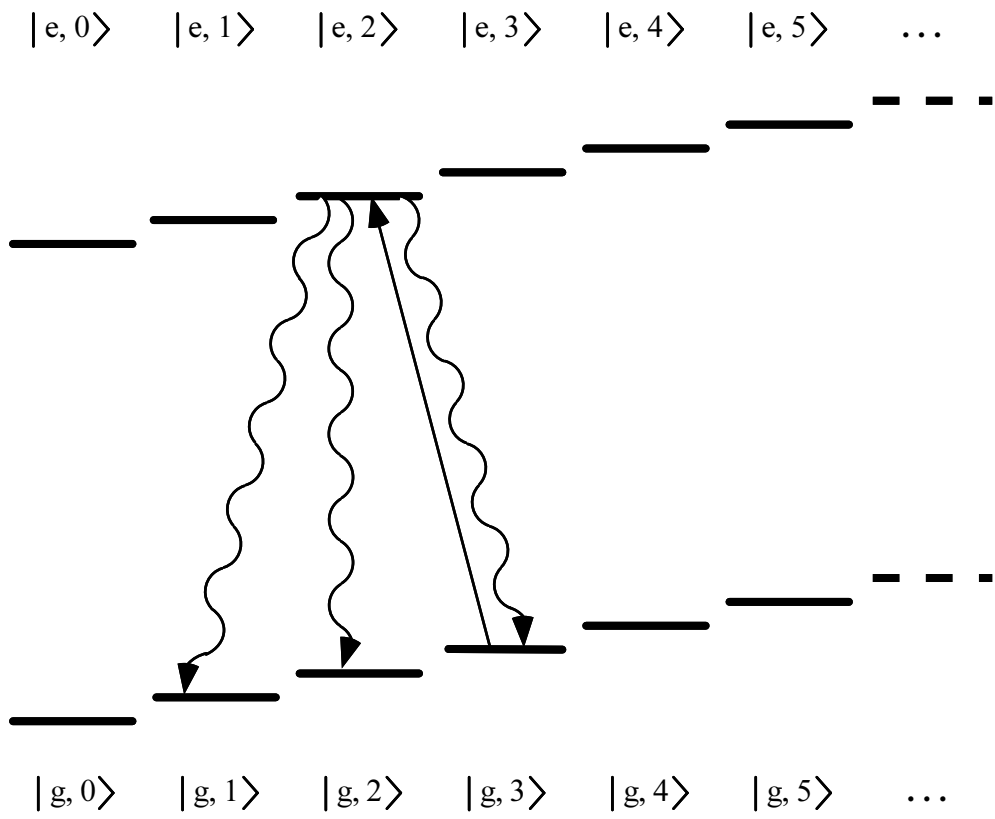
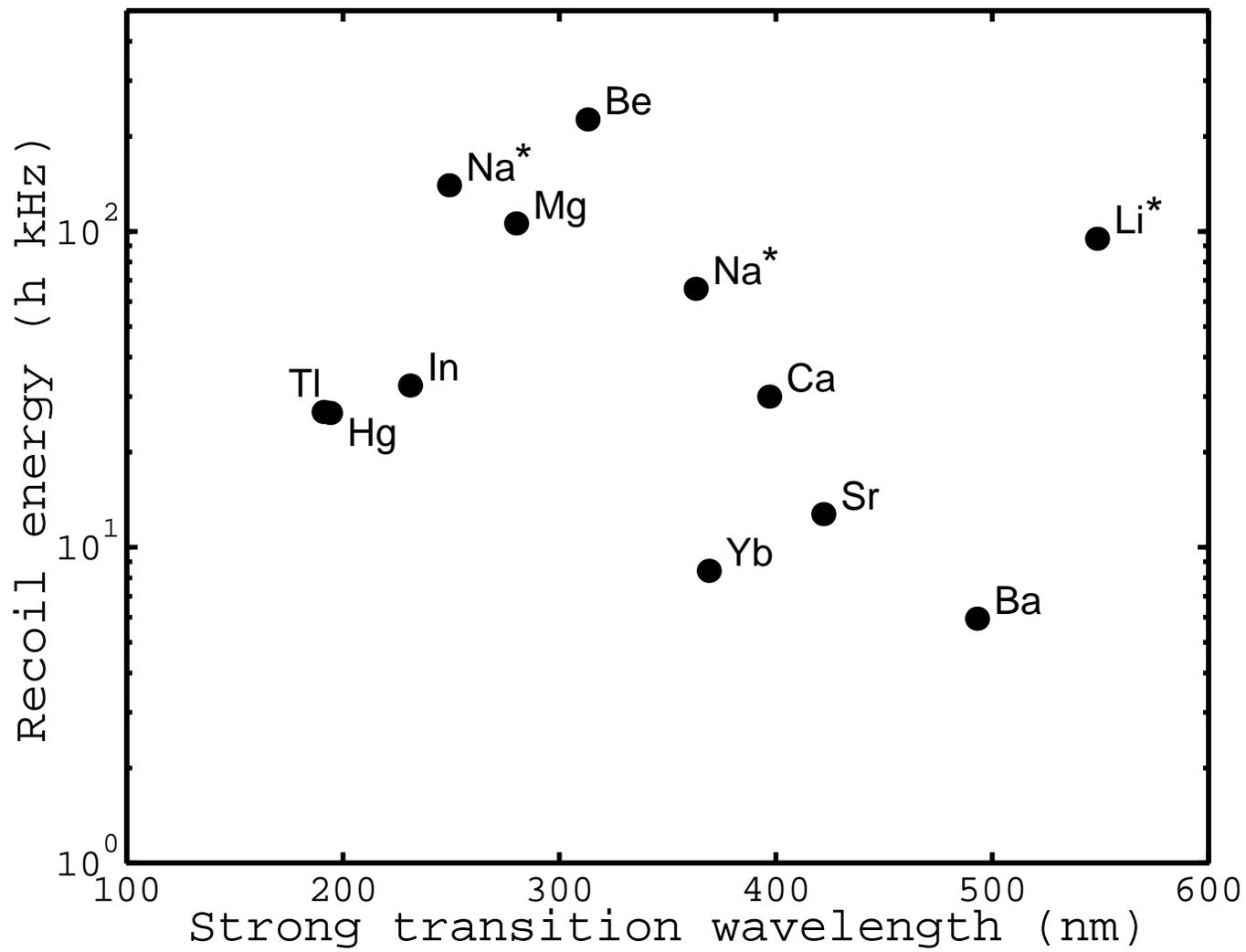
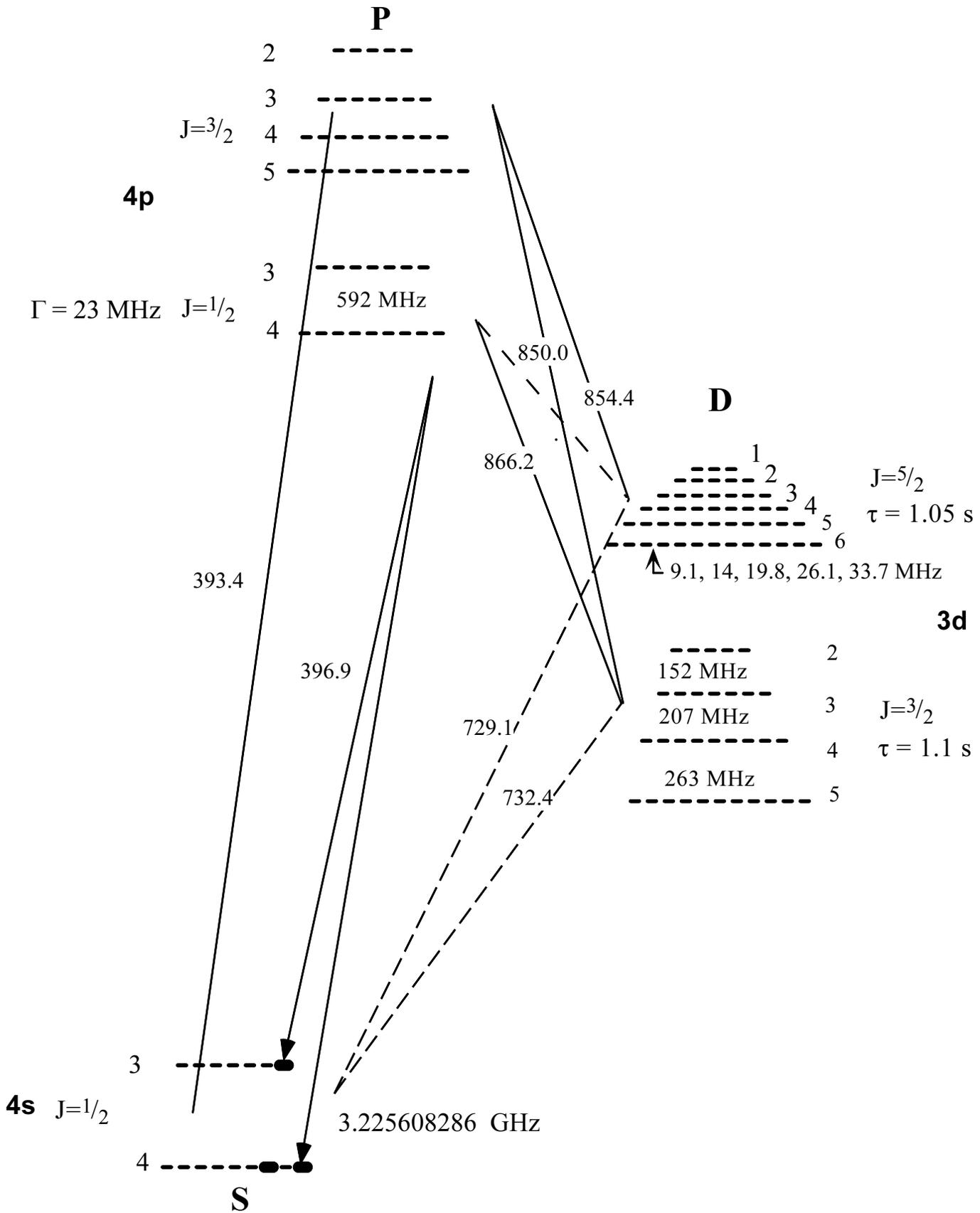
\end{document}